\documentclass[journal]{IEEEtran}

  \usepackage{fancyhdr}
  \usepackage{cite}
  \usepackage{url}
  \usepackage{amsmath}
  \usepackage{tabularx}
  \usepackage{subfigure}
  \usepackage{amsmath}
  \usepackage{breqn}

\ifCLASSINFOpdf
  \usepackage{graphicx}
\else
\fi

\usepackage{ifpdf}
\usepackage{graphicx}
\usepackage{amsmath}
\usepackage{pslatex}
\usepackage{xcolor}
\usepackage{textcomp}
\usepackage{placeins}
\usepackage{eqparbox}
\usepackage{url}
\usepackage{stfloats}
\usepackage{multicol}
\usepackage{lipsum}
\usepackage{float}
\usepackage{algorithmicx}
\usepackage{algpseudocode}
\usepackage{amsthm}
\usepackage{comment}
\usepackage{cite}
\usepackage{color, soul}
\usepackage{enumerate}
\usepackage{bbm}
\usepackage{threeparttable}
\usepackage{booktabs}
\usepackage{array}
\usepackage{balance}
\usepackage{color}
\usepackage{amssymb,amsmath}
\usepackage{capt-of}
\usepackage{xcolor}
\usepackage{amsbsy}

\begin{document}

\title{The Impact of Traffic Characteristics on System and User Performance in 5G/6G Cellular Systems}
    
\author{Eduard~Sopin, Vyacheslav Begishev, Vladislav Prosvirov, Konstantin Samuyilov, Yevgeni Koucheryavy
\vspace{-0mm}

\thanks{This work was supported by the Russian Science Foundation, project no. 23-79-01140 and by the grant for research centers in the field of AI provided by the Analytical Center for the Government of the Russian Federation (ACRF) in accordance with the agreement on the provision of subsidies (identifier of the agreement 000000D730321P5Q0002) and the agreement with HSE University No. 70-2021-00139.}

\thanks{Eduard~Sopin, Vyacheslav~Begishev, and Konstantin Samuyilov are with RUDN University, Moscow, Russia. Email:~\{sopin-es,begishev-vo, samuylov-ke\}@rudn.ru.}
\thanks{Vladislav Prosvirov and Yevgeni Koucheryavy are with Higher School of Economics, National Research University, Moscow, 101000 Russia Federation. E-mail: \{ykoucheryavy,vprosvirov@hse.ru\}}


\vspace{-0mm}
}

\maketitle

\begin{abstract}
The statistical characteristics of the propagation environment and traffic arrival process are known to affect the user performance in 5G/6G millimeter wave (mmWave) and sub-terahertz (sub-THz) systems. While the former topic has received considerable attention recently, little is known about the impact of traffic statistics. In this study, we characterize the effects of correlation and variability in the session arrival process on the performance of 5G/6G mmWave/sub-THz systems. To this end, we use the tools of stochastic geometry and queuing theory to model the service process at base stations (BS) and specifics of the mmWave/sub-THz radio part. The metrics considered include the system resource utilization and session loss probability. Our results show that the normalized autocorrelation function (NACF), coefficient of variation (CoV), and variance of the resource request distribution have a significant impact on the considered parameters. For the same arrival rate, high values of lag-1 NACF and CoV may lead the system out of the operational regime, affecting the loss probability and resource utilization by up to an order of magnitude. Even a slight deviation from the uncorrelated Poisson process decreases the utilization by 10-20\% and increases the session loss probability multiple times. Radio and environmental characteristics may further increase the variability in resource request distribution and decrease resource utilization. In general, the use of the commonly accepted Poisson assumption leads to a severe underestimation of the actual performance of 5G/6G mmWave/sub-THz systems. Therefore, both traffic arrival and propagation statistics are equally important for accurate performance assessment of such systems.
\end{abstract}



\section{Introduction}


The emergence of high capacity radio access technologies (RAT) that operate in the millimeter wave (mmWave) band, 30-100 GHz, and sub-terahertz (sub-THz) band, 100-300 GHz, is expected to enable the support of novel services at the air interface, such as high-resolution streaming, holographic communications, and extended virtual reality (x-VR) \cite{navarro2020survey,chowdhury20206g}. In contrast to modern applications, these services are rate-greedy in nature and require quality-of-service (QoS) guarantees in terms of bitrate stability \cite{dahlman20205g,giordani2020toward}.





User and system performance in 5G/6G mmWave/sub-THz systems are affected by the statistical features of the traffic arrival process and propagation environment. Specifically, the propagation in the mmWave/sub-THz bands is a complex phenomenon that has an impact on session service continuity. In addition to the large propagation losses, limiting the coverage of prospective mmWave/sub-THz base stations (BS), the blockage of propagation paths by various objects in the channel, such as human bodies, may lead to frequent loss of connectivity \cite{moltchanov2022tutorial}. As a result, performance evaluation models for 5G/6G mmWave/sub-THz cellular systems proposed in the past have mainly concentrated on capturing the propagation phenomenon (e.g., \cite{andrews2016modeling,bai2014coverage,kokkoniemi2017stochastic,petrov2015interference,shafie2021coverage}). 


\textcolor{black}{In contrast to propagation specifics, little is currently known about the impact of traffic arrival statistics in mmWave/sub-THz systems on user- and system-centric performance metrics. The unique feature of our study is that we account for burstiness of the session arrival process in cellular deployments. This is in contrast to conventional assumptions of the Poisson arrivals taken in many studies of cellular systems. Such type of burstiness naturally happens in any type of telecommunications network including cellular systems, where arrivals can be modulated by some external events \cite{qi2016characterizing,feng2018deeptp}, e.g., flash crowds, event-based services, social interactions of users within the cell coverage, etc.} 

\textcolor{black}{While the burniness of the session arrival process in cellular systems is confirmed in several studies, the authors are unaware of rigorous investigation of exact distributional and correlational properties in cellular systems. To this aim, in our study we concentrate on two main metrics affecting burstiness phenomenon, coefficient of variation (CoV) and normalized autocorrelation function (NACF).} It has long been known that these parameters of the session arrival process may have a negative impact on key performance indicators such as system resource utilization and session loss probability \cite{li1993queue,hajek1998variations}. Furthermore, it is known that the resource requirements of the sessions also depend on environmental factors such as propagation conditions and density of blockers \cite{moltchanov2022tutorial,kovalchukov2019accurate}. Understanding the impact of these factors is therefore critical for the deployment of mmWave/sub-THz 5G/6G systems.



\textcolor{black}{In our study, we consider the so-called “last mile”, i.e., the service process of users’ at the 5G/6G base station operating in mmWave/sub-THz band.} The aim of this paper is to characterize the impact of temporal correlation and variability in arrival traffic patterns on service performance in 5G/6G mmWave/sub-THz systems. To this end, we utilized stochastic geometry to capture the propagation properties and characterize the resource requirements of arriving sessions and queuing-theoretic tools to evaluate the delivered system- and user-oriented key performance indicators. In our numerical results, we determined the impact of the temporal correlation and variability of the arrival process on system resource utilization and session loss probability.


The contributions of our study include:
\begin{itemize}
    \item{mathematical model to account for the impact of the temporal correlation and variability of the session arrival process in 5G/6G mmWave/sub-THz systems with blockages and directional antennas;}
    \item{numerical results showing that variability and temporal clustering in the session arrival process negatively impact both system- and user-oriented performance indicators, decreasing resource utilization multiple times and increasing session loss probability;}
    \item{observation showing that the radio part and environmental parameters affect the system's performance mainly via the variability of the session's resource request distribution and thus need to be accurately captured to predict the actual system performance.}
\end{itemize}


The reminder of this paper is organized as follows. 
We review related work in Section \ref{sect:related}. The system model is presented in Section \ref{sect:syst}. The model is formalized and solved in Section \ref{sect:perf} and then parameterized in Section \ref{sect:parametr}. Numerical results demonstrating the impact of correlation and variability on user- and system-centric performance metrics are presented in Section \ref{sect:num}. Finally, the conclusions are provided in the last section.

\section{Related Work}\label{sect:related}

The choice of performance evaluation method for cellular systems depends heavily on the type of user traffic. Under elastic traffic assumptions, tools of stochastic geometry are conventionally utilized \cite{elsawy2016modeling,hmamouche2021new}. However, when non-elastic traffic, that is expected to be a typical load for 5G/6G systems, is considered, stochastic geometry needs to be complemented with queuing theory to capture the behavior of the session service process at BSs.


The performance of 5G/6G mmWave/sub-THz systems is often evaluated by assuming independence in both the spatial \cite{andrews2016modeling,petrov2017interference,shafie2020coverage} and temporal domains \cite{moltchanov2022tutorial,polese_dual_con}. To this end, most authors utilize the Poisson point process (PPP) to represent user locations in the service area of base stations (BS) and the Poisson arrival process of sessions from the users. Both the temporal and spatial versions of the Poisson process are intrinsically independent at all time instants and points, neglecting any possible correlation in these two domains. Additionally, the Poisson assumption limits the variability of the arrival process. Motivated by these facts, the authors of \cite{ghosh2020role} evaluated the impact of clustered user locations, highlighting that depending on the chosen user deployment model the performance metrics of the system can be as far as 10-30\% from those calculated by utilizing the PPP assumption. Similar conclusions have been observed in other studies that considered various deviations from PPP (e.g., \cite{turgut2019uplink,muhammad2020uplink}).

The authors in \cite{navarro2020survey} provided a detailed overview of traffic models utilized by standardization organizations, including both 3GPP and ITU-T and highlighted that most of them concentrated on the packet arrival process, neglecting session dynamics. However, it is known that the session arrival process in cellular systems may pose complex properties with occasional spikes in the offered traffic load \cite{yu2020step,zhao2020spatial,chen2015analyzing}. \textcolor{black}{Furthermore, studies performed in the past within queuing theory show that user- and system-centric metrics may vary dramatically \cite{li1993queue,hajek1998variations} in response to the NACF and CoV even when the mean session arrival intensity remains intact.}

\section{System Model}\label{sect:syst}

In this section, we introduce the system model by successively defining deployment, blockage, propagation, antenna, session arrival, and session service models. Finally, we define the metrics of interest.

\subsection{Deployment and Blockage}

\begin{figure}[!t]
\vspace{-0mm}
\centering
\includegraphics[width=1.0\columnwidth]{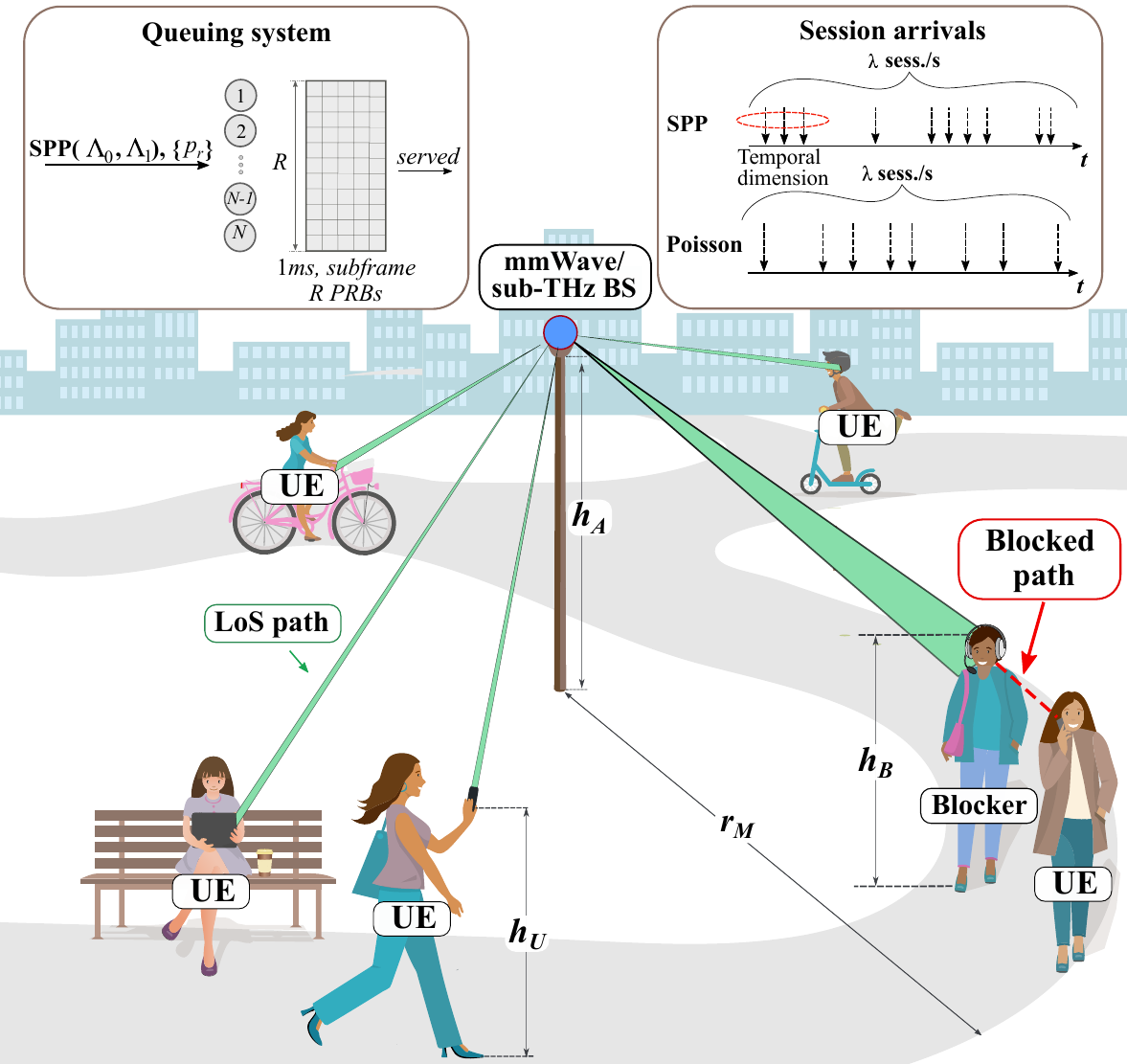}
\caption{Visual illustration of the considered system model's elements.}
\label{fig:deployment}
\end{figure}

In this study, we consider mmWave/sub-THz BS serving user equipment (UE) devices associated with pedestrians located in the coverage area of radius $r_C$. The cell coverage $r_C$ depends on the radio parameters and utilized modulation and coding schemes (MCS) and is obtained further in Section \ref{sect:perf}. The BS operates using a bandwidth $B$. The heights of the BS and UEs are constants, $h_A$ and $h_U$, respectively (see Fig. \ref{fig:deployment}). 



We assume that the considered area is populated with pedestrians producing human body blockage. The pedestrians move in the area by following the Random Direction Model (RDM) \cite{nain2005properties} with speed and mean run time parameters denoted by $v_B$ m/s and $\tau_B$ s, respectively. The flux of blockers through the cell boundary is assumed to be constant, that is, the blockers process in $\Re^2$ is stationary. The density of pedestrians is assumed to be $\lambda_B$ ped./km$^2$. 

We model blockers by cylinders with the following parameters: (i) base radius $r_B$ and, (ii) cylinder height $h_B$. Denoting by $r$ the two-dimensional (2D) distance between the UE and BS, we determine the pedestrian's body blockage probability, $p_B$, by utilizing the model from \cite{gapeyenko2016analysis}, that is,
\begin{align}\label{eqn:blockage}
p_B(r)=1-e^{-2\lambda_Br_B\left[r\frac{h_B-h_U}{h_A-h_U}+r_B\right]}.
\end{align}

\subsection{Propagation and Antenna Models}


To characterize the propagation properties, we utilize the Urban Micro (UMi) model standardized by 3GPP in TR 38.901 \cite{standard_16}. Specifically, the path losses at the three-dimensional (3D) distance $y$ are given by
\begin{align}\label{eqn:plDb}
L_{i}^{(dB)}(y)= 
\begin{cases}
(32.4 +\epsilon_0) + 10\zeta\log(y) + 20\log{f_{C}},\\
(32.4+\epsilon_1) + 10\zeta\log(y) + 20\log{f_{C}},\\
\end{cases}
\end{align}
where $i=0,1$ denotes the non-blocked and blocked states, respectively, $\epsilon_0=0$ dB and $\epsilon_1=15$ dB are the blockage attenuations, $f_C$ is the operational frequency measured in GHz, and $\zeta=2.1$ is the propagation exponent. 

This model can be written in the linear form as
\begin{align}\label{eqn:linearForm}
L_i(y)=
\begin{cases}
10^{2\log_{10}{f_{C}}+(3.24+\epsilon_0/10)}y^{-\zeta},\\
10^{2\log_{10}{f_{C}}+(3.24+\epsilon_1/10)}y^{-\zeta}.\\
\end{cases}
\end{align}

\textcolor{black}{We note that the model in (\ref{eqn:plDb}) and (\ref{eqn:linearForm}) is the 3GPP standardized UMi propagation model for 0.5-100 GHz. THz frequencies are also characterized by additional effects including atmospheric absorption \cite{boronin2014capacity}, potential molecular noise \cite{boronin2015molecular}, potential more severe scattering in typical urban deployments \cite{ometov2019packet}, and more diverse multi-path phenomenon, etc. However, the impact of the former factor in the sub-THz band is rather limited to selected frequencies, see Fig. A, and should not produce significant effects when one is to choose frequencies for sub-THz 6G cellular systems carefully (as it has been done for 5G mmWave systems avoiding 60 GHz band). Also, recall that the impact of molecular noise has been recently shown to be negligible \cite{kokkoniemi2016discussion}. Scattering off various objects in the channel such as foliage as well as other atmospheric effects (snow, fog, rain) may induce different attenuation at different frequencies even in the 0.5-100 GHz band. However, as shown in \cite{sen2022terahertz}, in the sub-THz band (120-140 GHz) the impact is comparable to mmWave band. The same applies to the multi-path effect.}

\textcolor{black}{Recently, a number of sub-THz propagation models have indeed been proposed, e.g., \cite{shurakov2023empirical,xing2021millimeter} and detailed summary in \cite{rappaport2022radio}. They demonstrated that for Friis-like models that are similar in structure to the 3GPP’s UMi model utilized in the paper, the joint impact of the abovementioned effects leads to different coefficients reducing the effective coverage of the cell and does not produce principally new propagation effects as compared to mmWave band. As the goal of our study is to assess whether traffic burstiness has an impact on the service performance in 5G/6G systems, and if yes, what is the magnitude of this impact, we decided to go with standardized 3GPP model.}



The BS and UEs are assumed to utilize planar antenna arrays. To represent the antenna radiation pattern, we utilize the cone model specified in \cite{interference1,petrov2017interference} capturing the half-power beamwidth (HPBW) of the array. By following \cite{constantine2005antenna}, the average gain of the HPBW is provided by 
\begin{align}\label{eqn:gain}
G = \frac{1}{\theta_{3db}^+-\theta_{3db}^{-}}\int_{\theta_{3db}^-}^{\theta_{3db}^+} \frac{\sin(N_{(\cdot)}\pi\cos(\theta)/2)}{\sin(\pi\cos(\theta)/2)}d\theta,
\end{align}
where the number of antenna elements in a specific plane (horizontal or vertical) is denoted by $N_{(\cdot)}$, and the HPBW given by $\theta_{3db}^+-\theta_{3db}^-$  is approximated by $102^\circ/N_{(\cdot)}$.

\subsection{Traffic and Service Processes}


Sessions are generated by the UEs associated with pedestrians. We consider a non-adaptive traffic with constant bitrate $C$. Note that owing to wireless channel impairments and the random location of UEs in the service area, the actual amount of resources associated with bitrate $C$ is random. 

Sessions arrive according to Markovian arrival process (MAP), which is known to have versatile distributional and correlational properties \cite{lucantoni1990map,chakravarthy2001batch}. The MAP is governed by a continuous-time Markov chain $\alpha(t)$ defined on the set of states $\mathcal{M}$, $|\mathcal{M}|=M$. The MAP flow is characterized by two matrices: $\pmb{\Lambda}_0$ and $\pmb{\Lambda}_1 $. The generator matrix of the underlying Markov chain is $\pmb{Q}=\pmb{\Lambda}_0+\pmb{\Lambda}_1$. 

Assuming that $\alpha (t)$ is irreducible, then $\vec{\theta}$ is its stationary distribution, and the average arrival rate $\lambda_A$ is given by
\begin{align}
\lambda_A=\vec{\theta} \pmb{\Lambda}_1 \vec{1},
\end{align}
where $\vec{1}$ is the unit vector of appropriate size.




Sessions are served in order or their arrivals. Upon arrival, a new session requests a server (i.e., a signal processor) and a random amount of resources represented by primary resource blocks (PRB). The number of requested PRBs follows the probability mass function (PMF) $p_{j}$, $1\leq{}j\leq{}R$, corresponding to the rate $C$, which is computed in Section \ref{sect:parametr}. When the system has sufficient amount of resources, the arriving session is accepted. Alternatively, the session is dropped. The service time of the accepted sessions follows an exponential distribution with a mean $\mu^{-1}$.



\subsection{\textcolor{black}{Metrics and Approach}}

In our study, we are interested in assessing the impact of the temporal correlation and variability of the session arrival process on system- and user-oriented key performance indicators, including system resource utilization and session loss probability. \textcolor{black}{We note that the resource utilization $U$ is equal to the fraction of the average number of occupied PRBs to the total number of available PRBs in the system.} 

\textcolor{black}{Although booth classes of metrics characterize the efficiency of the communications system’s performance, they have somewhat contradictory goals. From the user perspective, it is vital to get service that meets applications’ QoS requirements. In mmWave/sub-THz systems that are intended to serve heavy-bitrate, often non-elastic traffic, probability of session drop is one of the critical user-oriented metrics. While this metric is definitely of interest for network operators as low values of session drop probability improves user satisfaction of the service, network operators are also interested in utilizing resources as efficiently as possible as cellular operational bands are a large part of capital expenditures (CAPEX). To this aim, we consider the average number of PRBs representing resource utilization as another metric of interest.}

\textcolor{black}{To derive the metrics of interest, we take the following approach. First, in Section \ref{sect:perf}, by utilizing the introduced traffic and services processes we formulate a queuing model describing the service process of 5G/6G BS operating in mmWave/sub-THz band. Then, in Section \ref{sect:parametr}, we demonstrate how to derive the PMF $p_{j}$, $1\leq{}j\leq{}R$ of resources requested by a session accounting for the blockage, antenna, and propagation model introduced in this section.}

\section{Performance Evaluation}\label{sect:perf}

In this section, we first formalize and solve the queuing model describing the service process at mmWave/sub-THz BS.

\subsection{Session Service Model}


Consider a resource loss queuing system (ReLS) with $N$ servers, a finite set of resources consisting of $R$ PRBs, and MAP arrivals, where each arrival requests a random number of PRBs with PMF $\{p_{j}\}$. As shown in \cite{moltchanov2022tutorial}, because the resource requirements of the sessions are all random, the precise description of the system even with simplified Poisson arrivals requires a multi-dimensional Markov process. The rationale is that we need to track the amount of resources allocated to all sessions in the system. The use of MAP as the arrival process adds an additional degree of complexity.

We utilize a simplified approach to analyze the system. Accordingly, we keep track of only the total number of occupied resource units instead of a vector of occupied resources by all the sessions in the system. Although this helps reduce the state space of the considered system, one cannot determine the exact number of PRBs to be released upon a session departure. To address this issue, we utilize the Bayesian approach, that is, if there are $n$ sessions in the system that totally occupy $r$ PRBs, then one session occupies $j$ resource units with probability
\begin{align}
p_j p_{r-j}^{(n-1)}/p_r^{(n)},    
\end{align}
where $p_r^{(n)}$ is the probability that $n$ sessions together occupy $r$ PRBs, and can be evaluated as $n$-fold convolution of the initial distribution $\{p_j\}$. In \cite{naumov2016}, it was proven that under Poisson arrivals and exponential service times, the stationary distribution of the simplified system is equivalent to that of the initial system. Here, we utilize this as an approximate approach. 

The system behavior can be described by a three-dimensional Markov process 
\begin{align}
X(t)=(\xi(t), \delta(t), \eta(t)),
\end{align}
where $\xi(t)$ denotes the number of sessions currently in the system, $\delta(t)$ specifies the overall number of utilized PRBs, and $\eta(t)$ is the state of the underlying Markov chain. 

The system state space $\mathcal{S}$ can be defined as
\begin{align}
    &\mathcal{S}=\left( \{0\} \cup \bigcup_{k=1}^N \mathcal{S}_k \right) \times \mathcal{M}, \\
    &\mathcal{S}_k=\left\{ (k,r): 0 \leq r\leq R, p_r^{(k)}>0, \right\}, 1\leq k \leq N. \nonumber
\end{align}

Denote by $s_k$ the cardinality of $\mathcal{S}_k$. Let us arrange the elements $(k,r)$ in each of the subsets $\mathcal{S}_k$ in the ascending order of the second component. We then introduce an integer function $I(k,r)$, which is equal to the sequence number of $(k,r)$ in $\mathcal{S}_k$, if $p_r^{(k)}>0$, $r\leq R$, and $0$ otherwise.

Denote by $q_0(m)$ the stationary probability that there are no sessions in the system and the state of the underlying Markov chain is $m$, and let 
\begin{align}
\vec{q}_0=( q_0(1),\dots, q_0(M)).    
\end{align}

Similarly, denote by $q_{k,r}(m)$ the steady-state probability that $k$ sessions that are currently in the system occupy totally $r$ PRBs, and the state of the chain is $m$, that is,
\begin{align}
\vec{q}_k=( \vec{q}_{k,r_1}, \vec{q}_{k,r_2}\dots, \vec{q}_{k,r_{s_k}}),
\end{align}
where $\vec{q}_{k,r}=( q_{k,r}(1), \dots, q_{k,r}(M) )$. 

Since $X(t)$ is a quasi-birth-and-death (QBD, \cite{bean2010quasi}) process, the vectors $\vec{q}_0$ and $\vec{q}_{k,r}$ should satisfy the following system of equilibrium equations
\begin{align}\label{eqn:sur}
    &\vec{q}_0 \pmb{D}_0 + \vec{q}_1 \pmb{M}_1=\Vec{0}, \nonumber \\
    &\vec{q}_{k-1} \pmb{L}_{k-1} + \vec{q}_k \pmb{D}_k + \vec{q}_{k+1} \pmb{M}_{k+1}=\Vec{0}, \quad 1 \leq k \leq N,\\
    &\vec{q}_{N-1} \pmb{L}_{N-1} + \vec{q}_N \pmb{D}_N =\Vec{0}, \nonumber
\end{align}
where matrix $\pmb{D}_0$ is equal to $\pmb{\Lambda}_0$. 

The other matrices $\pmb{D}_k, 1 \leq k \leq N$ are square matrices with block-diagonal structures with $M \times M$ blocks. The dimensions of matrix $\pmb{D}_k$ are $s_k M \times s_k M$. The diagonal blocks of $\pmb{D}_k$ for $I(k,r)>0$ are given by
    \begin{align}
        &\pmb{d}_k(I(k,r),I(k,r))= \pmb{Q}-k\mu \pmb{I} - \left(\sum_{j=0}^{R-r} p_j \right) \pmb{\Lambda}_1, 1 \leq k < N,\nonumber \\
        &\pmb{d}_N(I(N,r),I(N,r))= \pmb{Q}-N\mu \pmb{I},
    \end{align}
where $\pmb{I}$ is the unit matrix of appropriate size.

The matrix $\pmb{L}_0$ is a row of $s_1$ blocks. The block sizes are also $M \times M$, and the block matrices of $\pmb{L}_0$ have the following form for all $I(1,r)>0$
\begin{align}
    \pmb{l}_0\left( I(1,r) \right)= p_r\pmb{\Lambda}_1.
\end{align}

The dimensions of the other matrices $\pmb{L}_k$, $1\leq k \leq N-1$ are $s_k M \times s_{k+1} M$, and their $M \times M$ blocks for all $I(k,r)>0$ and $I(k+1,s)>0$ are given by
\begin{align}
    \pmb{l}_k(I(k,r),I(k+1,s))= p_{s-r} \pmb{\Lambda}_1.
\end{align}

Finally, the matrix $\pmb{M}_1$ is a column of $s_1$ blocks. The block matrices of $\pmb{M}_1$ are diagonal for all $I(1,r)>0$, that is,
\begin{align}
    \pmb{m}_1(I(1,r)) = \mu \pmb{I}.
\end{align}

The dimensions of the other matrices $\pmb{M}_k$, $2\leq k \leq N$ are $s_k M \times s_{k-1} M$, and their $M \times M$ blocks for all $I(k,r)>0$ and $I(k-1,s)>0$ are given by
\begin{align}
    \pmb{m}_k(I(k,r),I(k-1,s)) = \frac{p_{r-s} p_s^{(k-1)}}{p_r^{(k)}} k\mu \pmb{I}.
\end{align}

\begin{figure}[t!]
\vspace{-0mm}
\centering
\includegraphics[width=1.0\columnwidth]{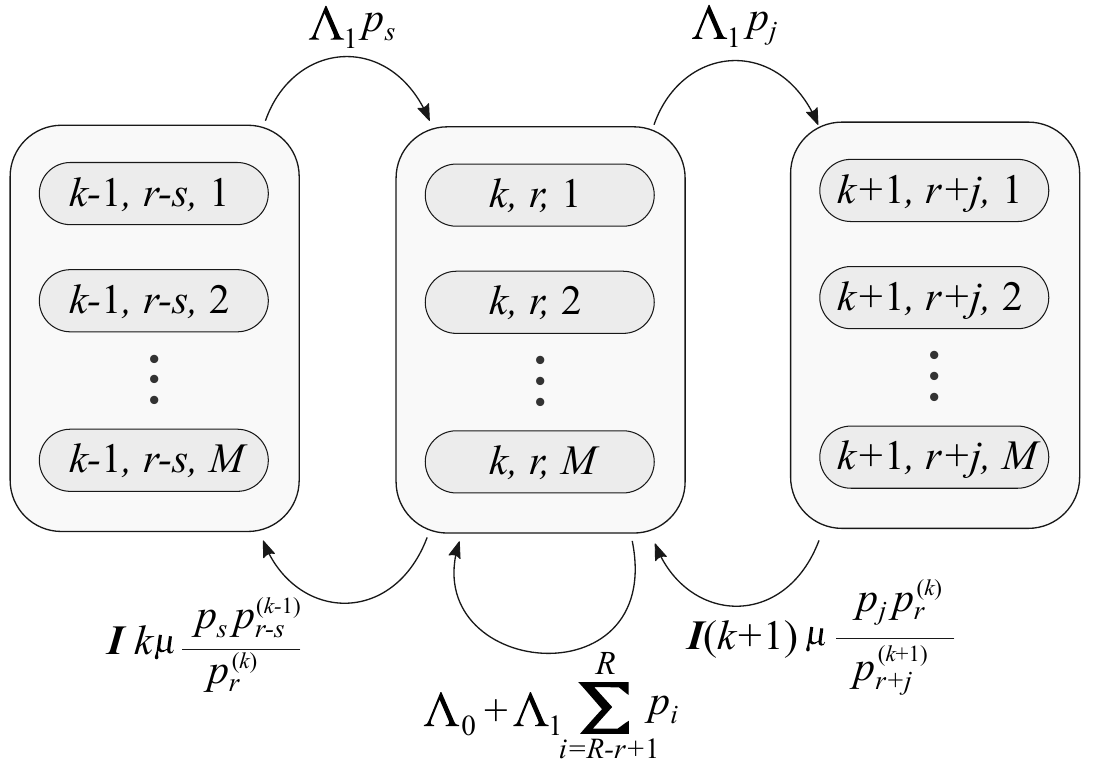}
\caption{A part of the state transition diagram of the model.}
\label{fig:state}
\end{figure}

\subsection{Equilibrium Equations and Solution}

The system in \eqref{eqn:sur} can be written in terms of the equidimensional vectors $\Vec{q}_0$, $\Vec{q}_{k,r}$, $1 \leq k \leq N$, $(k,r) \in \mathcal{S}_k$. The transition intensities caused by session arrivals are described by the matrix $\pmb{\Lambda}_1$. However, if the resource requirements of the session are greater than the unoccupied volume of system resources and, thus, there are no changes in the number of sessions in the system, then, the only possible changes are in the state of the underlying Markov chain. The matrix $\pmb{\Lambda}_0$ reflects the transition intensities of the underlying Markov chain without the arrival of a new session. Note that the departure of a session does not affect the state of the underlying Markov chain. A part of the state transition diagram with the corresponding intensities is illustrated in Fig. \ref{fig:state}.

By utilizing the abovementioned observations, the system in \eqref{eqn:sur} can be represented in the following form
\begin{align}\label{eqn:vecsur}
    &\vec{q}_0 \pmb{\Lambda}_0 + \mu \sum_{r=0}^R \vec{q}_{1,r}=\vec{0}, \nonumber \\
    &\sum_{j=0}^r p_j \vec{q}_{k-1,r-j} \pmb{\Lambda}_1 +\vec{q}_{k,r} \left( \pmb{Q}-k\mu\pmb{I}-\sum_{j=0}^{R-r} p_j \pmb{\Lambda}_1 \right) +  \\
    &+(k+1)\mu \sum_{j=0}^{R-r} \vec{q}_{k+1,r+j} \frac{p_j p_r^{(k)}}{p_{r+j}^{(k+1)}}=\vec{0}, 1 \leq k \leq N-1, \nonumber\\
    &\sum_{j=0}^r p_j \vec{q}_{N-1,r-j} \pmb{\Lambda}_1 +\vec{q}_{N,r} \left( \pmb{Q}-k\mu\pmb{I} \right)=\vec{0}. \nonumber
\end{align}

The representation in  \eqref{eqn:vecsur} makes the equilibrium equations more tractable. However, the block-tridiagonal structure of the corresponding generator matrix is still irregular. Nevertheless, system \eqref{eqn:vecsur} can be solved using any numerical method for systems of linear equations, including approximate iterative approaches such as Gauss-Seidel method \cite{milaszewicz1987improving}. Note that the equilibrium equations in \eqref{eqn:sur} can also be solved by numerical methods that can efficiently use the block-tridiagonal structure of the generator matrix, for example, \cite{akar1997finite}.

\subsection{Metrics of Interest}

We now proceed to determine the sought system- and user-oriented performance indicators, that is, the resource utilization and session loss probability. By utilizing the properties of the MAP arrival process, the session loss probability, $\pi_b$, is given by
\begin{align}
    \pi_b=1- \frac{1}{\lambda_A} \left( \sum_{k=0}^{N-1} \sum_{r: (k,r) \in \mathcal{S}_k} \vec{q}_{k,r} \sum_{j=1}^{R-r} p_j \right)  \pmb{\Lambda}_1 \Vec{1}.
\end{align}

\textcolor{black}{The resource utilization $U$ is equal to the fraction of the average number of occupied PRBs to the total number of available PRBs in the system. that is,
\begin{align}
    U=\frac{1}{R}\sum_{k=1}^{N} \sum_{r: (k,r) \in \mathcal{S}_k} \sum_{m \in \mathcal{M}} r q_{k,r}(m).
\end{align} }

\begin{table}[b!]
\vspace{-0mm}
\begin{center}
\caption{The mapping between channel quality indicators.}
\label{table:MCS}
\setlength{\tabcolsep}{3pt}
\begin{tabular}{llllll}\toprule
  CQI & Modulation & Level & Code rate & Spectral eff. & SINR\\ \hline\hline
   0  &   outage   &       &                  &&          \\ \hline
   1  &    QPSK    &   2   &      78/1024     &0.152& --9.478   \\ \hline
   2  &    QPSK    &   2   &     120/1024     &0.234& --6.658   \\ \hline
   3  &    QPSK    &   2   &     193/1024     &0.377& --4.098   \\ \hline
   4  &    QPSK    &   2   &     308/1024     &0.602& --1.798   \\ \hline
   5  &    QPSK    &   2   &     449/1024     &0.877&  0.399   \\ \hline
   6  &    QPSK    &   2   &     602/1024     &1.176&  2.424   \\ \hline
   7  &   16QAM    &   4   &     378/1024     &1.477&  4.489   \\ \hline
   8  &   16QAM    &   4   &     490/1024     &1.914&  6.367   \\ \hline
   9  &   16QAM    &   4   &     616/1024     &2.406&  8.456   \\ \hline
  10  &   64QAM    &   6   &     466/1024     &2.730& 10.266   \\ \hline
  11  &   64QAM    &   6   &     567/1024     &3.322& 12.218   \\ \hline
  12  &   64QAM    &   6   &     666/1024     &3.902& 14.122   \\ \hline
  13  &   64QAM    &   6   &     772/1024     &4.523& 15.849   \\ \hline
  14  &   64QAM    &   6   &     873/1024     &5.115& 17.786   \\ \hline
  15  &   64QAM    &   6   &     948/1024     &5.555& 19.809   \\ \bottomrule
\end{tabular}
\end{center}
\vspace{-0mm}
\end{table}

\section{Model Parameterization}\label{sect:parametr}

We now proceed demonstrating how to parameterize the system introduced and solved in the previous section by expressing the resource requirements of an arriving session as a function of radio and environmental parameters.

To parameterize the model, we need to provide PMF of resources requested by an arriving session, $\{p_{j}\}$. This parameter depends on the radio part and environmental parameters and can be obtained as demonstrated below. In what follows, we first determine the effective cell coverage $r_C$ defined as the maximum distance from the BS, where the UE does not experience an outage in blocked conditions. We then proceed deriving PMF $\{p_{j}\}$.

Denoting by $0$ and $1$ the non-blocked and blocked states, respectively, the value of the signal-to-interference plus noise ratio (SINR) at the UE can be approximated by the following weighted sum
\begin{align}\label{eqn:prop_mmWave}
S(y)=(1-p_B)\frac{W_0\Omega_0\Upsilon_0}{N_{T}B+I}y^{-\zeta}+p_B\frac{W_1\Omega_1\Upsilon_1}{N_{T}B+I}y^{-\zeta},
\end{align}
where $W_i=P_{T}G_{A}G_{U}L_i(y)$, $P_{T}$ is the radiated BS power, $G_{A}$ and $G_{U}$ represent the gains at the BS and UE sides, respectively, $I$ denotes the interference, $N_T$ is the thermal noise, $\Upsilon_i$ and $\Omega_i$ are the fast and shadow fading in blocked and non-blocked states, $y$ is the 3D distance between the UE and BS, $B$ is the bandwidth \cite{standard_16}, and $p_B$ is the spatially-averaged blockage probability.

To simplify the calculation, in what follows, we use interference, fast, and shadow fading margins, that is, $I_M$, $\Upsilon_{M,i}$, and $\Omega_{M,i}$, instead of the actual values of the corresponding parameters, as is often done in practice. We note that the latter two margins are provided in the 3GPP TR 38.901 recommendation \cite{standard_16}, whereas the interference margin can be calculated by utilizing the models in \cite{petrov2017interference,shafie2021coverage}. Thus, (\ref{eqn:prop_mmWave}) can be written as
\begin{align}\label{eqn:prop_margins}
S(y)&=(1-p_B)\frac{W_0}{N_{T}B+I_M+\Upsilon_{M,0}+\Omega_{M,0}}y^{-\zeta}+\nonumber\\
&+p_B\frac{W_1}{N_{T}B+I_M+\Upsilon_{M,1}+\Omega_{M,1}}y^{-\zeta}.
\end{align}

The effective coverage radius $r_C$ of a BS can now be determined based on the outage threshold, $S_{\min}$, when the UE is in the blockage state at the cell edge. By utilizing the second branch in (\ref{eqn:prop_margins}) and solving it with respect to the distance, we obtain
\begin{align}\label{eqn:coverage}
r_{C}=\left(\frac{W_1y^{-\zeta}}{(N_TB+I_M+\Omega_{M,1}+\Upsilon_{M,1})S_{\min}}
\right)^{1/\zeta}.
\end{align}

The SINR cumulative distribution function (CDF) in the non-blocked state for a coverage radius computed in (\ref{eqn:coverage}) is then given by
\begin{align}\label{F_S_Bn(s)}
F_{S,0}(s) = \frac{Q^2 - \left(\frac{A_0}{s}\right)^{\frac{2}{\zeta}}}{r_C^2},\,\frac{A_0}{Q^{\zeta}} \leq s < \frac{A_0}{(h_{A}-h_U)^{\zeta}}.
\end{align}
where the shortcuts are
\begin{align}
&Q=\sqrt{r_C^2+(h_{A}-h_U)^2},\nonumber\\
&A_0=\frac{W_0}{N_TB+I_M+\Omega_{M,1}+\Upsilon_{M,1}}.
\end{align} 

The SINR CDF in the blocked state, $F_{S,1}(s)$, can be obtained similarly while the final SINR CDF, $F_{S}(s)$, is obtained by weighting the resulting CDFs with the spatially averaged blockage probability $p_B$ computed using (\ref{eqn:blockage}) and $r_C$ as
\begin{align}
p_B=\int_{r_B}^{r_C}p_B(r)dr,
\end{align}
where $r_B$ is the blocker's radius.

Finally, $\{p_{j}\}$ is obtained by discretizing $F_{S}(s)$ according to the SINR to MCS mapping as discussed in \cite{nrmcs}. It is noteworthy that this mapping is proprietary. In what follows, we utilize the mapping provided in \cite{kovalchukov2019accurate} and presented in Table \ref{table:MCS}.

\section{Numerical Results}\label{sect:num}

In this section, we present our results. \textcolor{black}{To assess the impact of temporal correlation and variability of the session arrival process on user- and system-centric performance, we utilize a special case of MAP -- a switched Poisson process (SPP).} For this process, the parameters are directly controllable, as demonstrated below. The system parameters utilized in the numerical analysis are summarized in Table \ref{tab:parTable}. \textcolor{black}{We note that the performance evaluation model allows treating general MAP. However, their parameterization is very complex \cite{klemm2003modeling,casale2011building} and needed when one is sure about precise distributional properties of the arrival process that is rarely known in practice.}

\begin{table}[t!]\footnotesize
\vspace{-0mm}
\centering
\caption{System parameters for numerical study}
\label{tab:parTable}
\begin{tabular}{p{5.0cm}p{2.9cm}}
\hline
\textbf{Parameter}&\textbf{Value} \\
\hline\hline
Operational frequency, $f_c$ 			 					& 28\,GHz\\
\hline
Height of BS, $h_{A}$ 		 						& 10\,m\\
\hline
UE height, $h_{U}$ 								& 1.5\,m\\
\hline
Bandwidth, $B$ 										& 100\,MHz\\
\hline
Blocker height, $h_{B}$							& 1.7\,m\\
\hline
Velocity of the blocker, $v_B$ 									& 1\,m/s\\
\hline
Blocker radius, $r_B$ 									& 0.4\,m\\
\hline
Mean blocker run time, $1/\tau_B$ 			& 5\,m\\
\hline
Outage threshold, $S_{\min}$ 						& -9.47\,dB\\
\hline
Transmit power, $P_T$ 								& 2\,W\\
\hline
Path loss exponent, $\zeta$ 							& 2.1\\
\hline
Interference margin, $I_M$ 	& 3 dB\\
\hline
Fast and shadow fading margins, $\Upsilon_M$, $\Omega_M$ 	& 3 dB\\
\hline
Blocker intensity, $\lambda_B$ 						& 0.04\,units/m$^2$\\
\hline
Thermal noise, $N_T$ 									& -174\,dBm/Hz\\
\hline
Blockage attenuation, $\epsilon_1$ 							& 15\,dB\\
\hline
UE antenna array & $4\times{}4$\,el.\\
\hline
Requested session rate, $C$ 							& 10\,Mbps\\
\hline
Mean session service time, $1/\mu$ 					& 30\,s\\
\hline
Session arrival intensity, $\lambda$ 					& 0.1, sess./s\\
\hline
BS antenna array	 						& $16\times{}16$\,el.\\
\hline
\end{tabular}
\vspace{-0mm}
\end{table}

\subsection{Switched Poisson Process}

Let $\alpha(t)$ be a Markov process with two states, $\mathcal{M}=\{1,2\}$. The arrival process can then be completely parameterized using the following matrices
\begin{align}
\pmb{Q}=
\begin{pmatrix}
-r_1 & r_1 \\
r_2 & -r_2
\end{pmatrix},\,
\pmb{\Lambda}_1=
\begin{pmatrix}
\lambda_1 & 0 \\
0 & \lambda_2
\end{pmatrix},\,
\end{align}
where $\pmb{Q}$ is the infinitesimal generator and $\pmb{\Lambda}_1$ is the rate matrix with Poisson arrival rates associated with each state. It should be noted that $\pmb{\Lambda}_0$ can be obtained as $\pmb{\Lambda}_0 = \pmb{Q} - \pmb{\Lambda}_1$. This process is referred to as the SPP.

The steady-state probability vector for the Markov chain embedded at the arrival epochs $\vec{\theta}^*$ is given by \cite{fischer1993markov}
\begin{align}
\vec{\theta}^*(-\pmb{\Lambda}_0)^{-1}\pmb{\Lambda}_1=\vec{\theta}^*,
\end{align}
leading to the following explicit result for SPP
\begin{align}
\vec{\theta}^*=\left(\frac{\lambda_1r_2}{\lambda_1r_2+\lambda_2r_1},\frac{\lambda_2r_1}{\lambda_1r_2+\lambda_2r_1}\right).
\end{align}

By utilizing the joint Laplace-Stiltjes transform of the successive interarrival times, $X_1,X_2,\dots,X_n$, \cite{fischer1993markov}
\begin{align}
L(s_1,s_2,\dots,s_n)=\vec{\theta}^*\prod_{k=1}^{n}((s_k\pmb{I}-\pmb{\Lambda}_0)^{-1}\pmb{\Lambda}_1)\vec{1},
\end{align}
$\vec{1}$ is the unit vector, one may derive the CDF of successive interarrival times of SPP in the following form
\begin{align}\label{eqn:spp_cdf}
F(x)&=1-\vec{\theta}^*e^{-\pmb{\Lambda}_0x}(-\pmb{\Lambda}_0)^{-1}\pmb{\Lambda}_1\vec{e}=\nonumber\\
&=1-qe^{-u_1x}+(1-q)e^{-u_2x},
\end{align}
where $u_1<u_2$ and
\begin{align}
&u_1=\frac{\lambda_1+\lambda_2+r_1+r_2-\delta}{2},\nonumber\\
&u_2=\frac{\lambda_1+\lambda_2+r_1+r_2+\delta}{2},\nonumber\\
&q=\frac{\lambda_2^2r_1+\lambda_1^2r_2}{(\lambda_2r_1+\lambda_1r_2)(u_1-u_2)}-\frac{u_2}{u_1-u_2},
\end{align}
where $\delta$ is provided by
\begin{align}
\delta=\sqrt{(\lambda_1-\lambda_2+r_1-r_2)^2+4r_1r_2}.
\end{align}

Note that (\ref{eqn:spp_cdf}) is the CDF of the hyperexponential distribution with two phases having a CoV greater than or equal to one \cite{iversen2010teletraffic}.

By utilizing (\ref{eqn:spp_cdf}), the mean interarrival time can be expressed as
\begin{align}
E[X]=\frac{1}{\lambda_A}=\frac{(1-q)u_1+qu_2}{u_1u_2}=\frac{r_1+r_2}{\lambda_1r_2+\lambda_2r_1}.
\end{align}

Also, the autocovariance function is known to be \cite{fischer1993markov}
\begin{align}\label{eqn:ACF_gen}
K(i)&=E[(X_1-E[X_1])(X_{i+1}-E[X_{i+1}])]=\sigma_A^{2}\beta_A,
\end{align}
where the $\sigma_A^{2}$ is the variance provided by
\begin{align}
\sigma_A^{2}=\frac{(\lambda_1-\lambda_2)^{2}r_1r_2}{(\lambda_2r_1+\lambda_1r_2)^2(\lambda_1\lambda_2+\lambda_2r_1+\lambda_1r_2)},
\end{align}
while $\beta_A$ is the non-unit eigenvalue of the stochastic matrix $(-\pmb{\Lambda}_0)^{-1}\pmb{\Lambda}_1$ expressed as
\begin{align}\label{eqn:beta_nacf}
\beta_A=\frac{\lambda_1\lambda_2}{\lambda_1\lambda_2+\lambda_2r_1+\lambda_1r_2}.
\end{align}

Observing the structure of (\ref{eqn:ACF_gen}) one can recognize $\beta_A$ as the lag-1 value of the NACF.

Having three parameters that can be measured from the statistical data, mean, variance, and lag-1 NACF value, $(\lambda_A,\sigma_A^{2},\beta_A)$, one may observe that we need four, $(\lambda_1,\lambda_2,r_1,r_2)$ to completely parameterize the model. By choosing the arrival rate $\lambda_2$ as a free parameter such that $\lambda_2>\lambda_A=1/E[X]$, the remaining parameters can be estimated as
\begin{align}
\lambda_1&=\frac{\beta_A  (\lambda_2 E[X] -1)}{\beta_A  E[X]  (\lambda_2 E[X] -1)+\lambda_2 \sigma_A^2},\nonumber\\
r_1&=-\frac{(\beta_A -1) \lambda_2^2 \sigma_A^2  (\lambda_2 E[X] -1)}{(\beta_A  E[X]  (\lambda_2 E[X] -1)+\lambda_2 \sigma_A^2 )}\times{},\nonumber\\
&\times{}\frac{1}{\left(\beta_A  (\lambda_2 E[X] -1)^2+\lambda_2^2 \sigma_A^2 \right)}\nonumber\\
r_2&=-\frac{(\beta_A -1) \lambda_2 (\lambda_2 E[X] -1)^2}{\beta_A  (\lambda_2 E[X] -1)^2+\lambda_2^2 \sigma_A^2 }.
\end{align}

\subsection{Main Performance Insights}

\begin{figure}[!t]
\vspace{-0mm}
\centering\hspace{-0mm}
\subfigure[{Session loss probability}]{
    \includegraphics[width=0.45\textwidth]{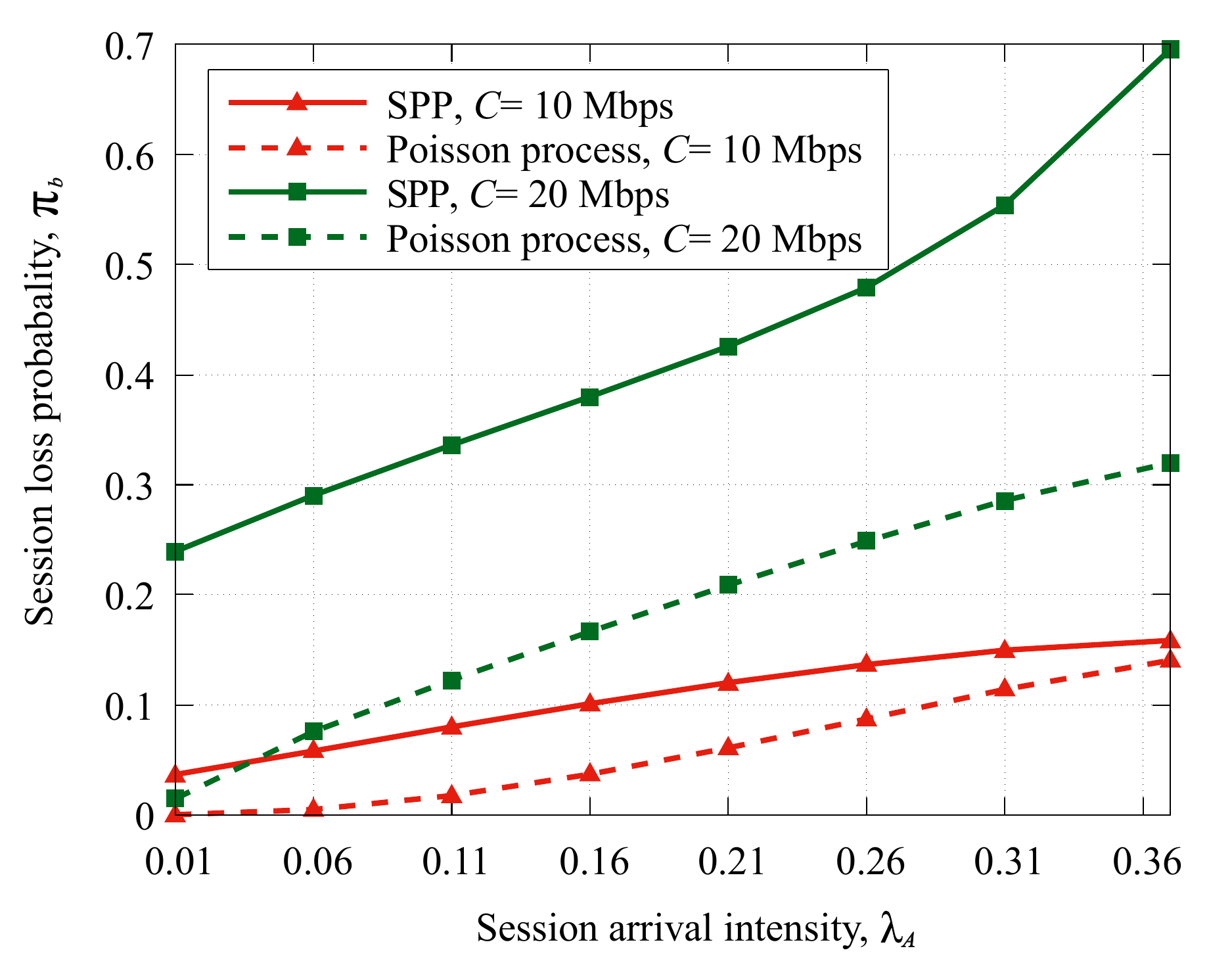}
    \label{fig:lambda_A_loss}
}\vspace{-0mm}\\
\subfigure[{System resource utilization}]{
    \includegraphics[width=0.45\textwidth]{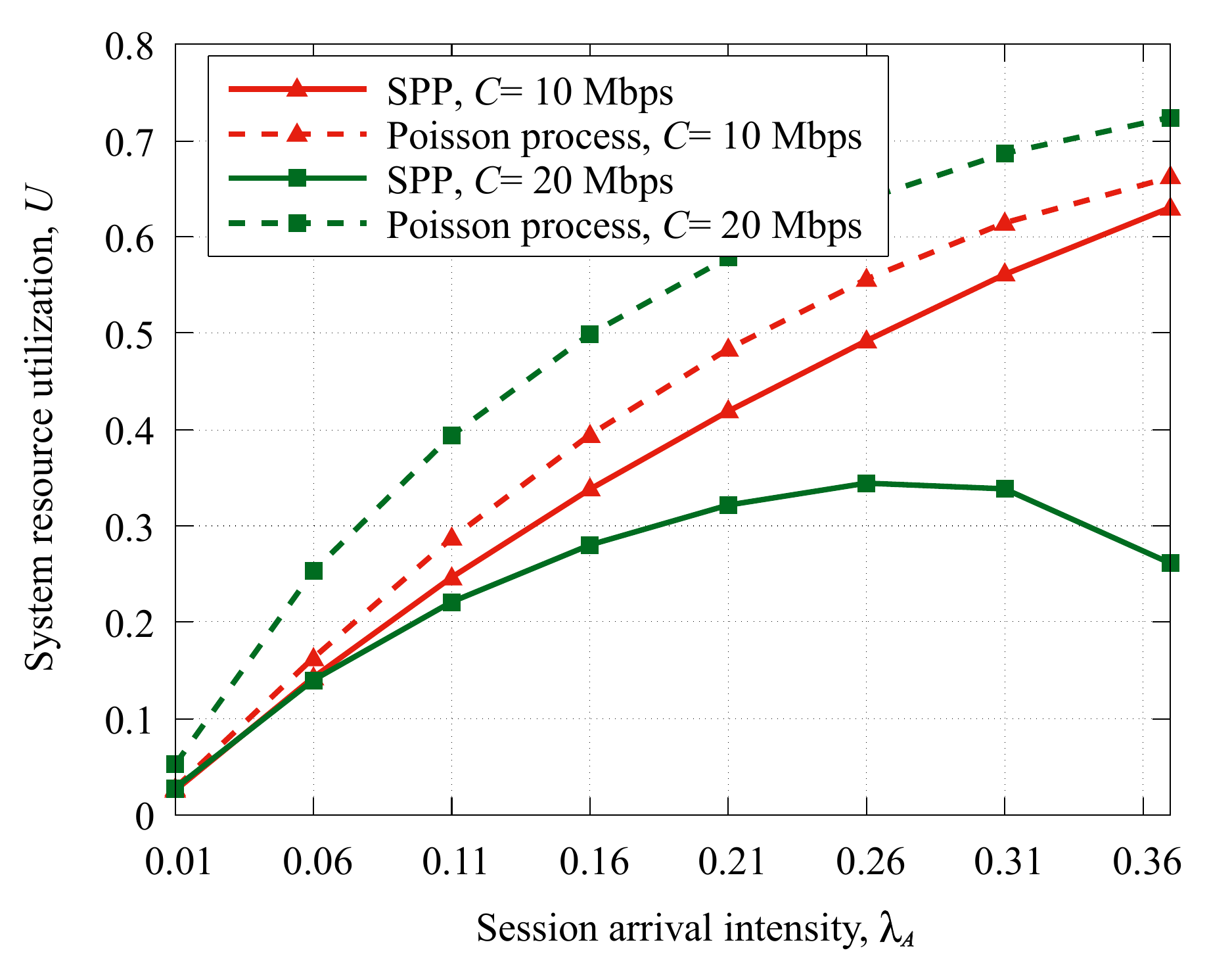}
    \label{fig:lambda_A_util}
}
\vspace{-0mm}
\caption{Performance metrics of interest as a function of arrival rate.}
\label{fig:lambda_A}
\vspace{-0mm}
\end{figure}


We start our numerical analysis in Fig. \ref{fig:lambda_A} by demonstrating the impact of the session arrival rate $\lambda_A$ on the session loss probability and system resource utilization for two different session rates of $C=10,20$ Mbps, lag-1 NACF of $\beta_A=0.1$, BS antenna array of $16\times{}16$ elements, and mean session service time of 30 s. Note that for all the considered values of $\lambda_A$ in Fig. \ref{fig:lambda_A}, we choose the variance of the session arrival rate, $\sigma_A^2$, such that the CoV $c_A=\sigma_A^2/\lambda_A$ is constant and equal to 2. For comparison purposes, here, and in what follows, we also show the results for the uncorrelated Poisson arrival process with the same arrival rate $\lambda_A$. Recall, that the CoV for the Poisson process is always one.


\textcolor{black}{By analyzing the data presented in Fig. \ref{fig:lambda_A}, we observe that both the session loss probability and system resource utilization are significantly affected by the probabilistic characteristics of the arrival traffic patterns.} Specifically, the SPP process, characterized by a higher variability than the Poisson one, is associated with drastically worse performance. The magnitude of the impact is also heavily affected by the session rate $C$. Specifically, for $C=20$ Mbps and the considered SPP process characterized by lag-1 NACF correlation $\beta_A=0.1$ and CoV $c_A=2$, the difference in the session loss probability is almost constant across the considered range of session arrival rates. For example, when $\lambda_A=0.03$ sess./s, the Poisson arrival process results in just approximately $5$\% lost sessions. However, when utilizing SPP, variability in the session arrival process leads the system out of the operational regime resulting in around $25$\% of lost sessions. Thus, the difference in terms of session loss probability can reach 200-400\% in absolute values.

\begin{figure}[!b]
\vspace{-0mm}
\centering
\includegraphics[width=0.45\textwidth]{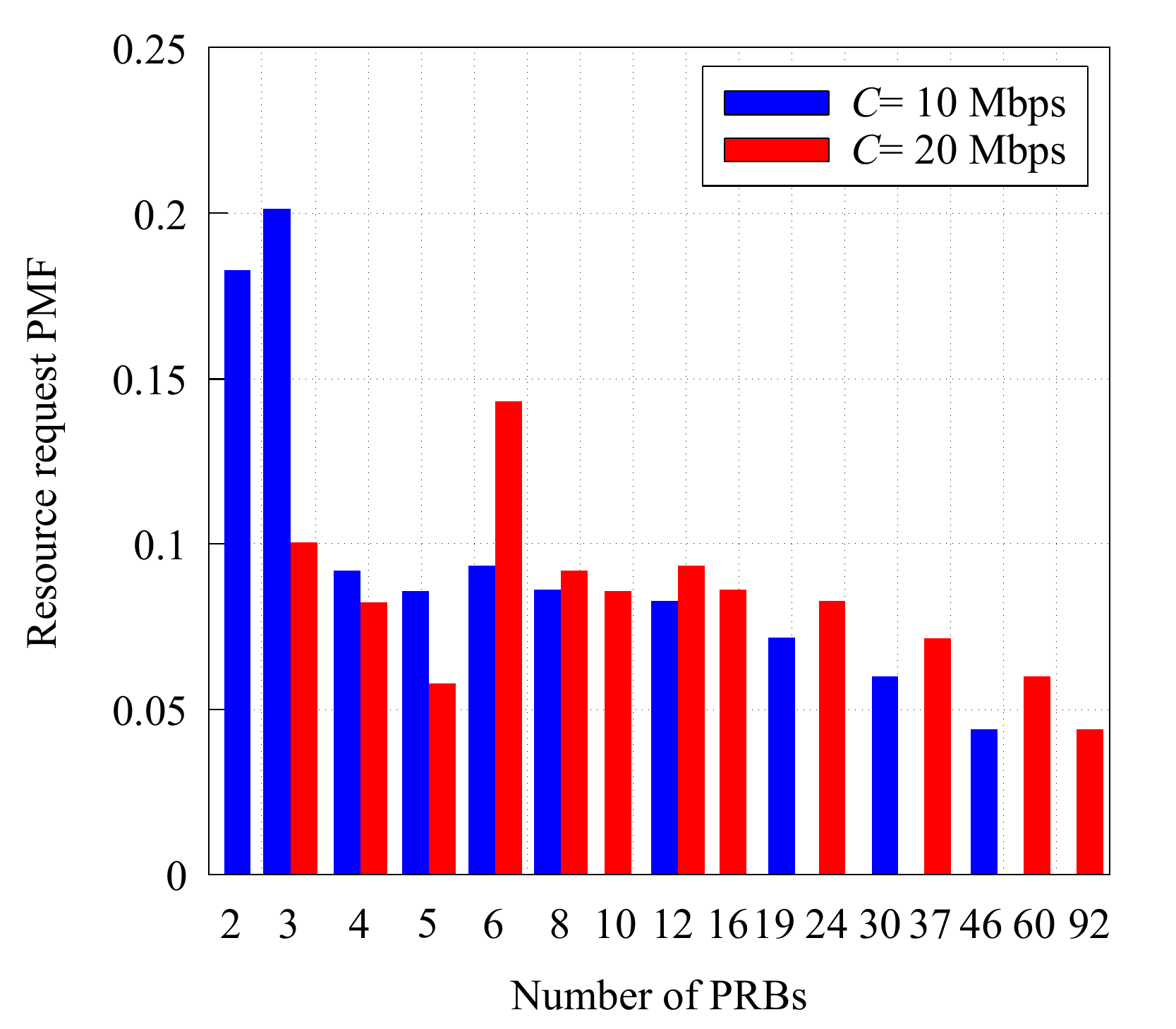}
\caption{PMFs of the session's resource requirements.}
\label{fig:res_pmfs}
\end{figure}


Analyzing the system resource utilization shown in Fig. \ref{fig:lambda_A_util}, we observe that Poisson traffic is associated with significantly better resource utilization than SPP traffic. Specifically, even for a session rate of $C=10$ Mbps the resource utilization induced by Poisson arrivals is higher than that of the SPP process with $C=20$ Mbps for the same arrival rate. On top of this, as one may observe, the SPP process with $C=20$ Mbps is characterized by the decreased utilization for high arrival rates. The difference increases as $\lambda_A$ increases and can reach $30-40$\%. These effects are explained by the impact of the resource request PMF associated with sessions with high session rate requirements, as demonstrated in Fig. \ref{fig:res_pmfs} for 10 Mbps and 20 Mbps. It can be observed that, the PMF corresponding to $C=20$ Mbps has several probabilities corresponding to extremely large values of the requested PRBs. In practice, these probabilities are related to the UEs located at the cell edge, and thus experiencing low SINR conditions. Under high arrival rates, i.e., $\lambda_A>0.25$ sess./s, sessions from these UEs are dropped because of an insufficient amount of available PRBs in the system upon arrival. This leads to increased session loss probabilities evident in Fig. \ref{fig:lambda_A_loss} and decrease in system resource utilization, as shown in Fig. \ref{fig:lambda_A_util}.

\begin{figure}[!t]
\vspace{-0mm}
\centering\hspace{-0mm}
\subfigure[{Session loss probability}]{
    \includegraphics[width=0.45\textwidth]{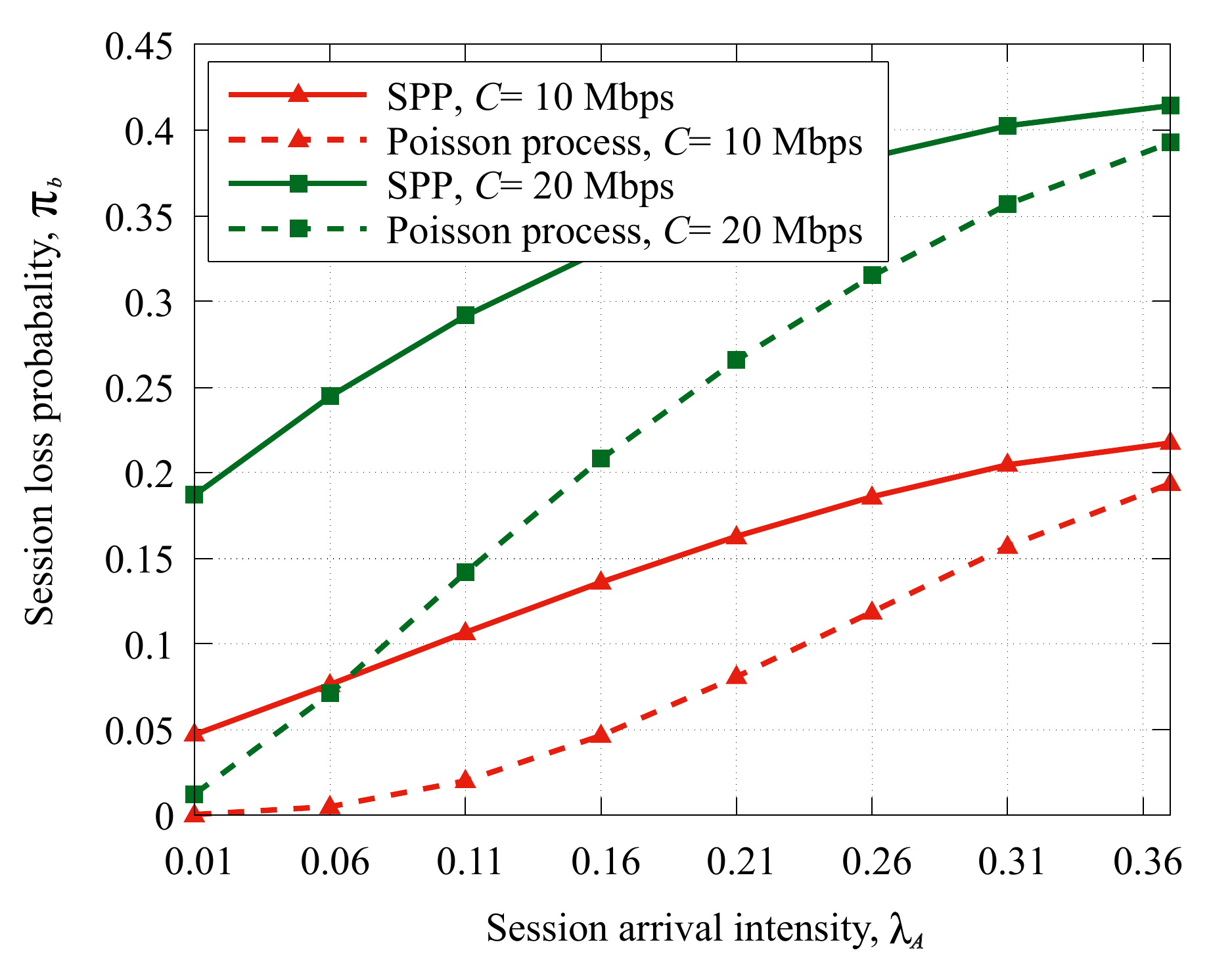}
    \label{fig:lambda_A_theory_loss}
}\vspace{-0mm}\\
\subfigure[{System resource utilization}]{
    \includegraphics[width=0.45\textwidth]{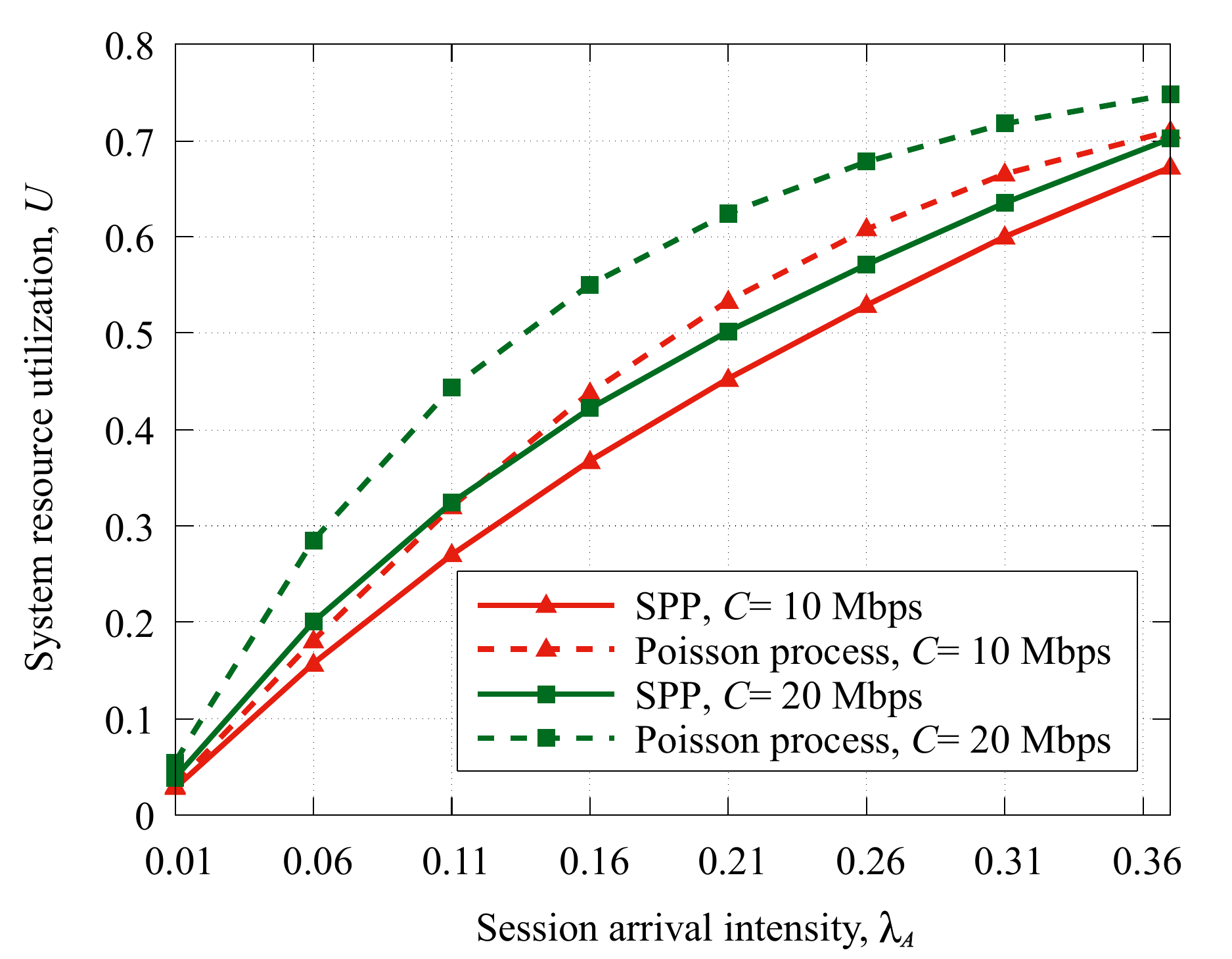}
    \label{fig:lambda_A_theory_util}
}
\vspace{-0mm}
\caption{Performance metrics for geometrically distributed PMF.}
\label{fig:lambda_A_theory}
\vspace{-0mm}
\end{figure}


To support our conclusions, in Fig. \ref{fig:lambda_A_theory} we show the session loss probability and system resource utilization for geometrically distributed session resource requirements with the mean coinciding with the mean of the actual resource request PMFs corresponding to the rates $C=10,20$ Mbps. The remaining parameters are the same as those in Fig. \ref{fig:lambda_A}. As can be observed, the trends related to the impact of lag-1 NACF and the CoV are preserved, that is, the SPP process is always characterized by much worse performance as compared to the Poisson process with the same mean. However, as the variance of the geometric distribution is much smaller (variance is $\lambda_A$ resulting in a CoV $c_A=1/\sqrt{\lambda_A}$, where $\lambda_A$ is the arrival intensity), we see that even for higher arrival intensities, there is no increase in session loss probability and a decrease in system resource utilization for $C=20$ Mbps.

\textcolor{black}{Note that the curve in Fig. \ref{fig:lambda_A_util} corresponding to SPP with $C=20$ Mbps has an inflection point. The rationale is that for $C=20$ Mbps, in the distribution of resource requirements there is 0.05 probability that a single incoming session requires 92\% of the system's resources. With Poisson arrivals, these sessions are almost always lost. However, in a SPP flow with the increased burstiness, this may not always be the case. In between traffic bursts, the system may become almost empty. And at the beginning of the next burst, a session demanding 92\% of the system's resources can be accepted for service. As a result, the system is unable to handle most of the sessions during this burst of traffic. Therefore, the increased burstiness leads to a reduction in session “discrimination” with respect to their resource requirements, but it also reduces system performance. The decrease of the system utilization $U$ at high loads with $C = 20$ Mbps is caused by the same effect.  However, we note that this is rather rare effect and only happens for very high data rates and rather limited systems’ resources.}

\subsection{Impact of NACF and CoV in Detail}

\begin{figure}[!t]
\vspace{-0mm}
\centering\hspace{-0mm}
\subfigure[{Session loss probability}]{
    \includegraphics[width=0.45\textwidth]{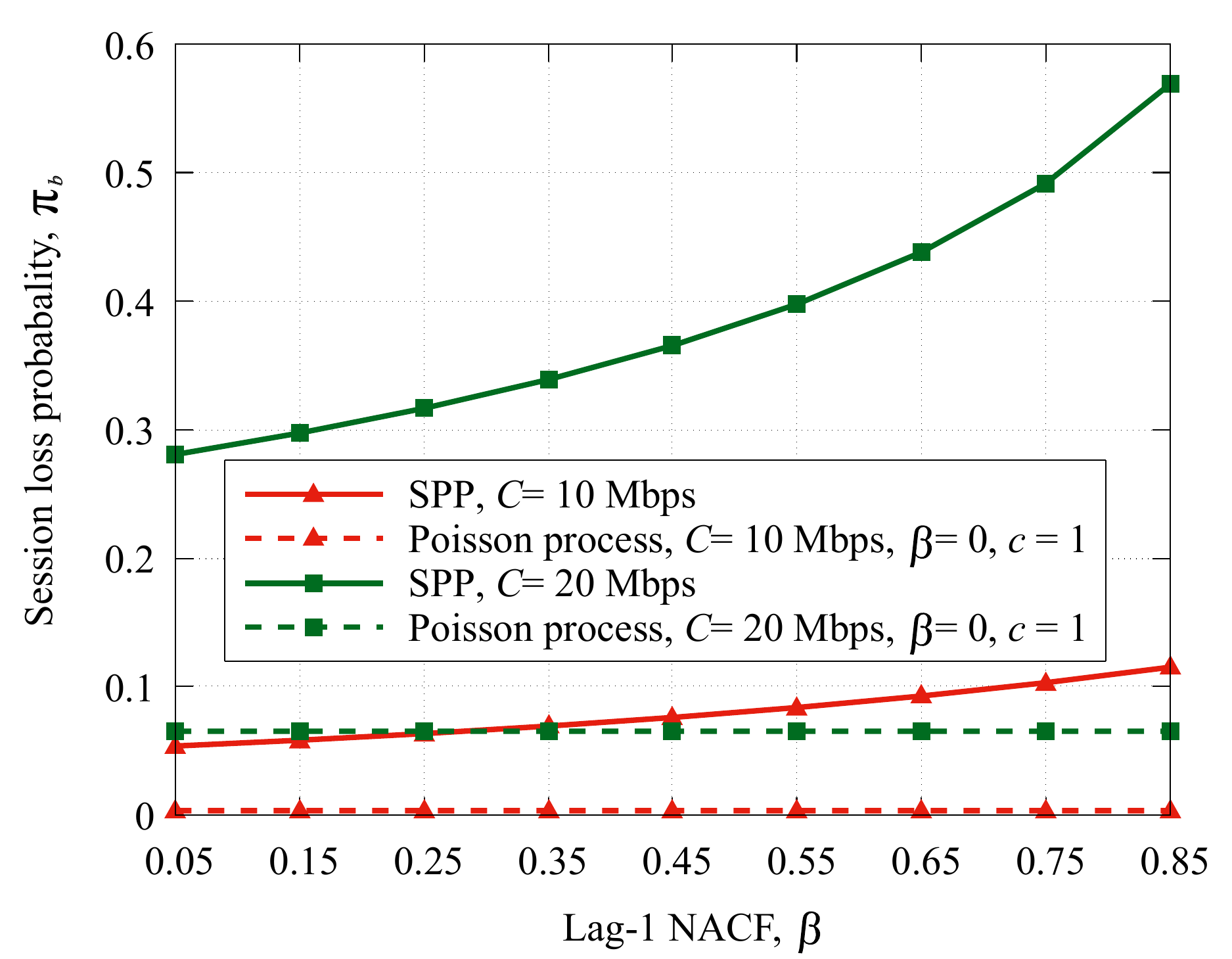}
    \label{fig:beta_loss}
}\vspace{-0mm}\\
\subfigure[{System resource utilization}]{
    \includegraphics[width=0.45\textwidth]{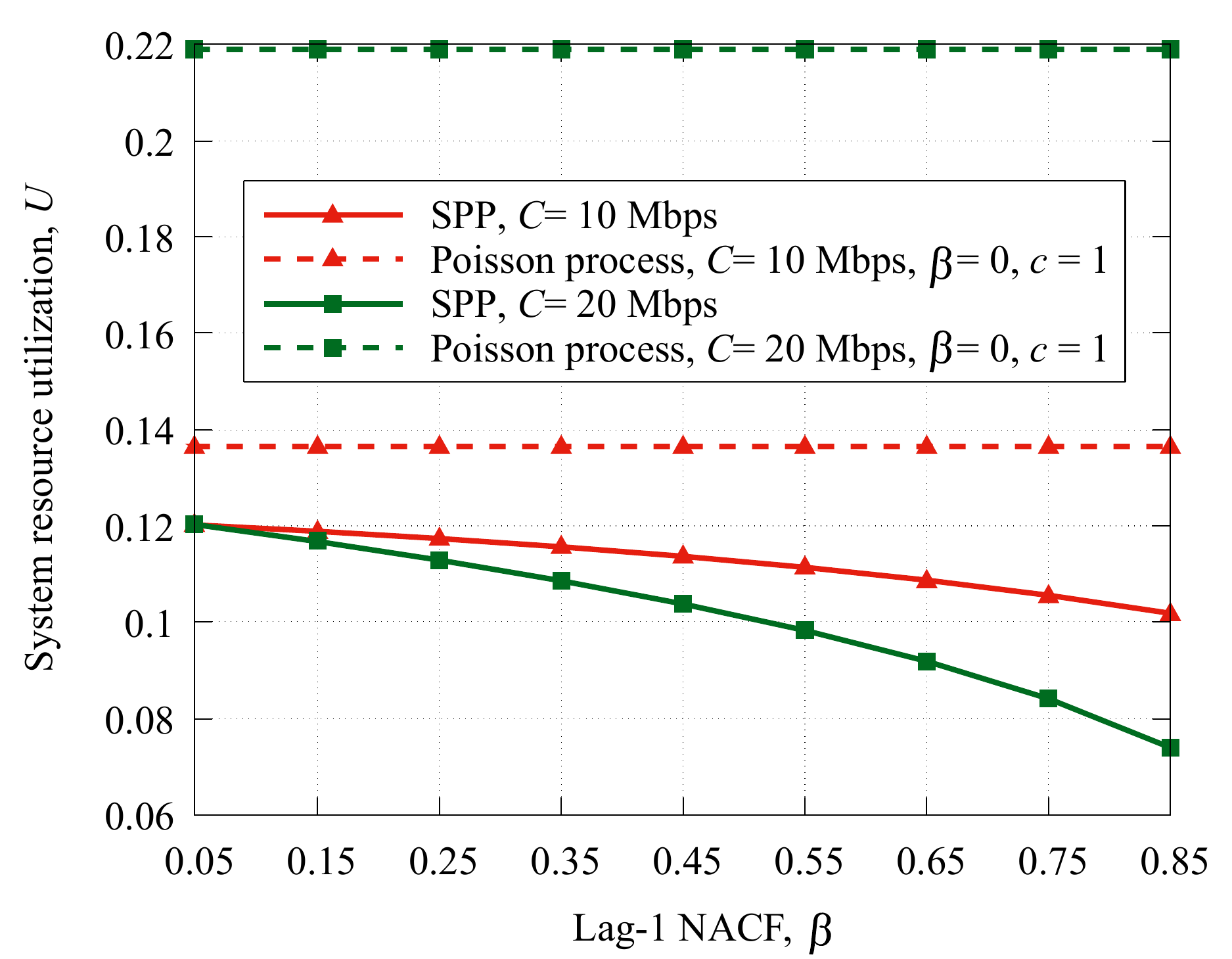}
    \label{fig:beta_util}
}
\vspace{-0mm}
\caption{Performance metrics as a function of lag-1 NACF.}
\label{fig:beta}
\vspace{-0mm}
\end{figure}


The results shown in Fig. \ref{fig:lambda_A} highlight that the impact of the NACF and the CoV can be large even for small values of $\beta_A$ and $c_A$. We will now study the magnitude of their impact in detail. To this end, Fig. \ref{fig:beta} shows the impact of the lag-1 NACF on the session loss probability and system resource utilization for two different session rates of $C=10,20$ Mbps, CoV $c_A=2$, BS antenna array of $16\times{}16$ elements, mean session service time of 30 s, and session arrival rate of $\lambda_A=0.1$ sess./s. 

By analyzing the results presented in Fig. \ref{fig:beta}, it can be observed that the impact of lag-1 NACF is smaller than that of the CoV for both metrics. Specifically, for $C=10$ the difference between the session loss probabilities corresponding to $\beta_A=0.1$ and $\beta_A=0.8$ is just $0.06$. For a higher session rate of $C=20$ Mbps, the session loss probability increases from $0.3$ to $0.6$ when the lag-1 NACF increases from $\beta_A=0.1$ to $\beta_A=0.8$. In both cases, the increase is two times, while the difference between the Poisson process and SPP process for $\beta_A=0.1$ is already larger than three times. Qualitatively similar observations can be made observing the system resource utilization shown in Fig. \ref{fig:beta_util}. 

\begin{figure}[!t]
\vspace{-0mm}
\centering\hspace{-0mm}
\subfigure[{Session loss probability}]{
    \includegraphics[width=0.45\textwidth]{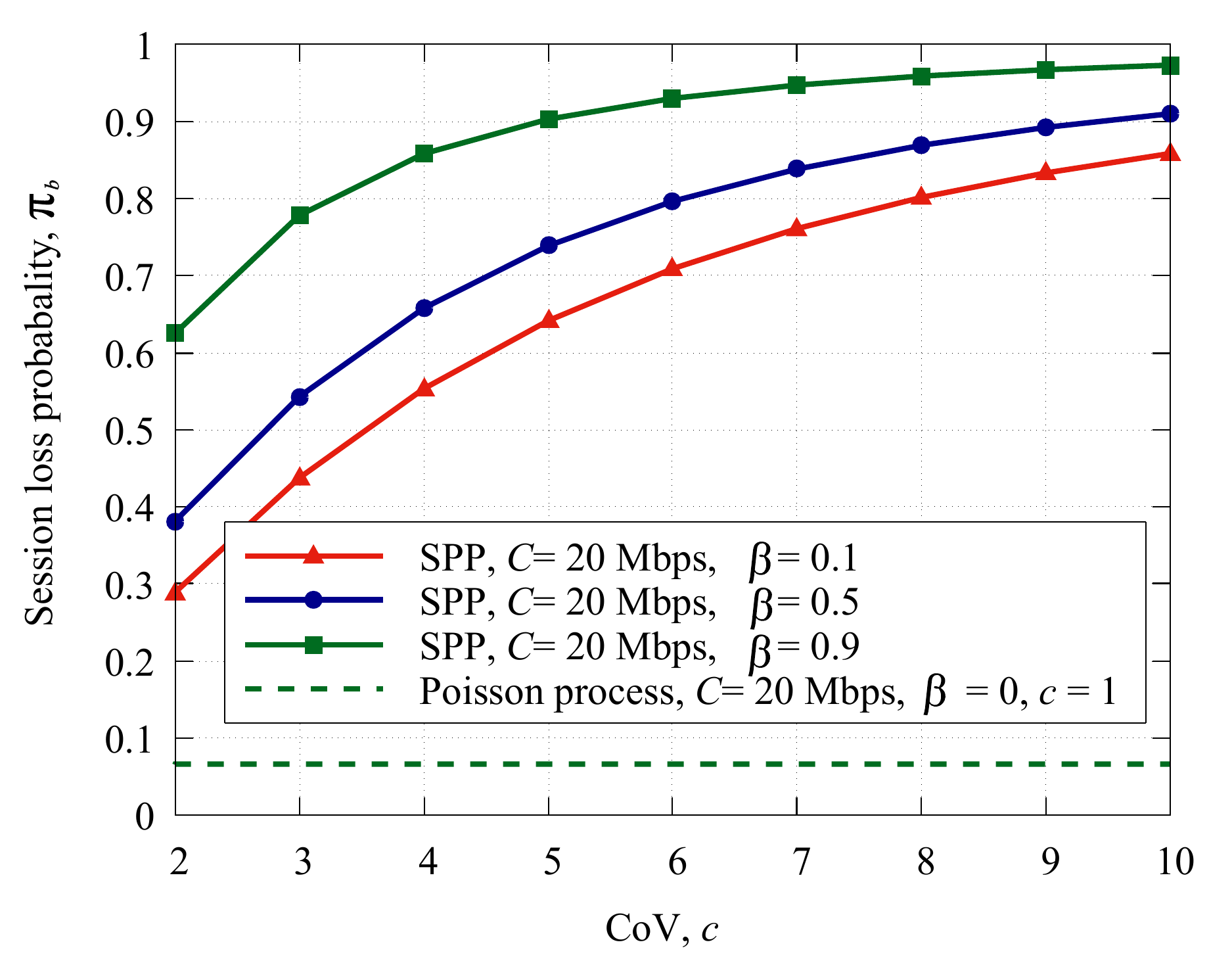}
    \label{fig:cov_loss}
}\vspace{-0mm}\\
\subfigure[{System resource utilization}]{
    \includegraphics[width=0.45\textwidth]{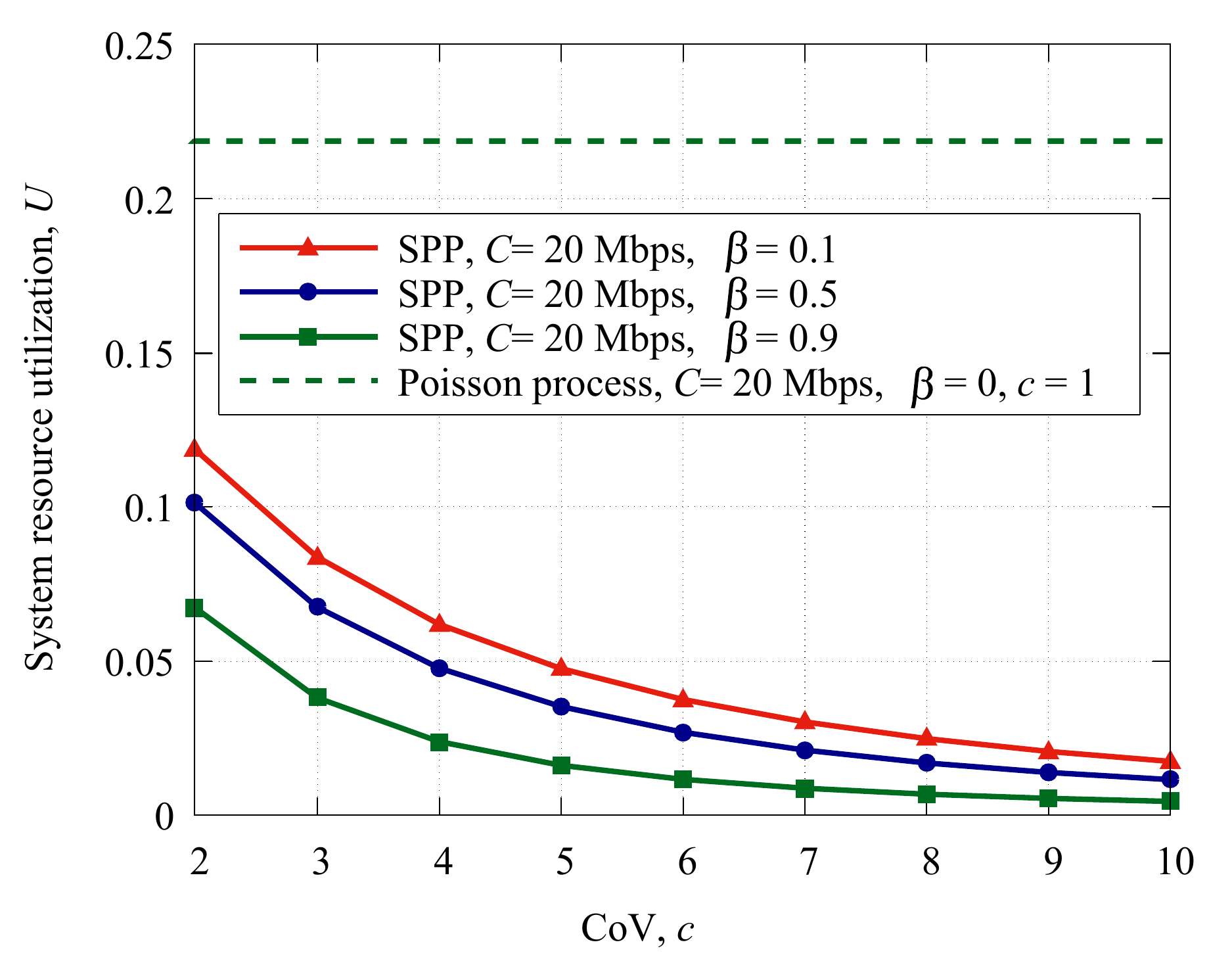}
    \label{fig:cov_util}
}
\vspace{-0mm}
\caption{Performance metrics as a function of CoV.}
\label{fig:cov}
\vspace{-0mm}
\end{figure}


The impact of the CoV of the session arrival process on the session loss probability and system resource utilization is further shown in Fig. \ref{fig:cov} for three different values of lag-1 NACF $\beta_A=0.1,0.5,0.9$, session rate of $C=20$ Mbps, BS antenna array of $16\times{}16$ elements, mean session service time of 30 s, and session arrival rate of $\lambda_A=0.1$ sess./s. It can be observed that, quantitatively, an increase in the CoV significantly affects both considered metrics. Large values of $c_A$ produce an order of magnitude increase in the session loss probability by leading the system out of the operational regime. The CoV has the greatest impact for low values of lag-1 NACF. Specifically, for completely uncorrelated arrivals with $c_A=1$, the system is characterized by non-negligible loss probability, whereas for $c_A=10$ it reaches one. Similar observations can be made regarding the system resource utilization shown in Fig. \ref{fig:cov_util}, where for $c_A=10$ it drops to zero.

\begin{figure}[!t]
\vspace{-0mm}
\centering\hspace{-0mm}
\subfigure[{Session loss probability}]{
    \includegraphics[width=0.45\textwidth]{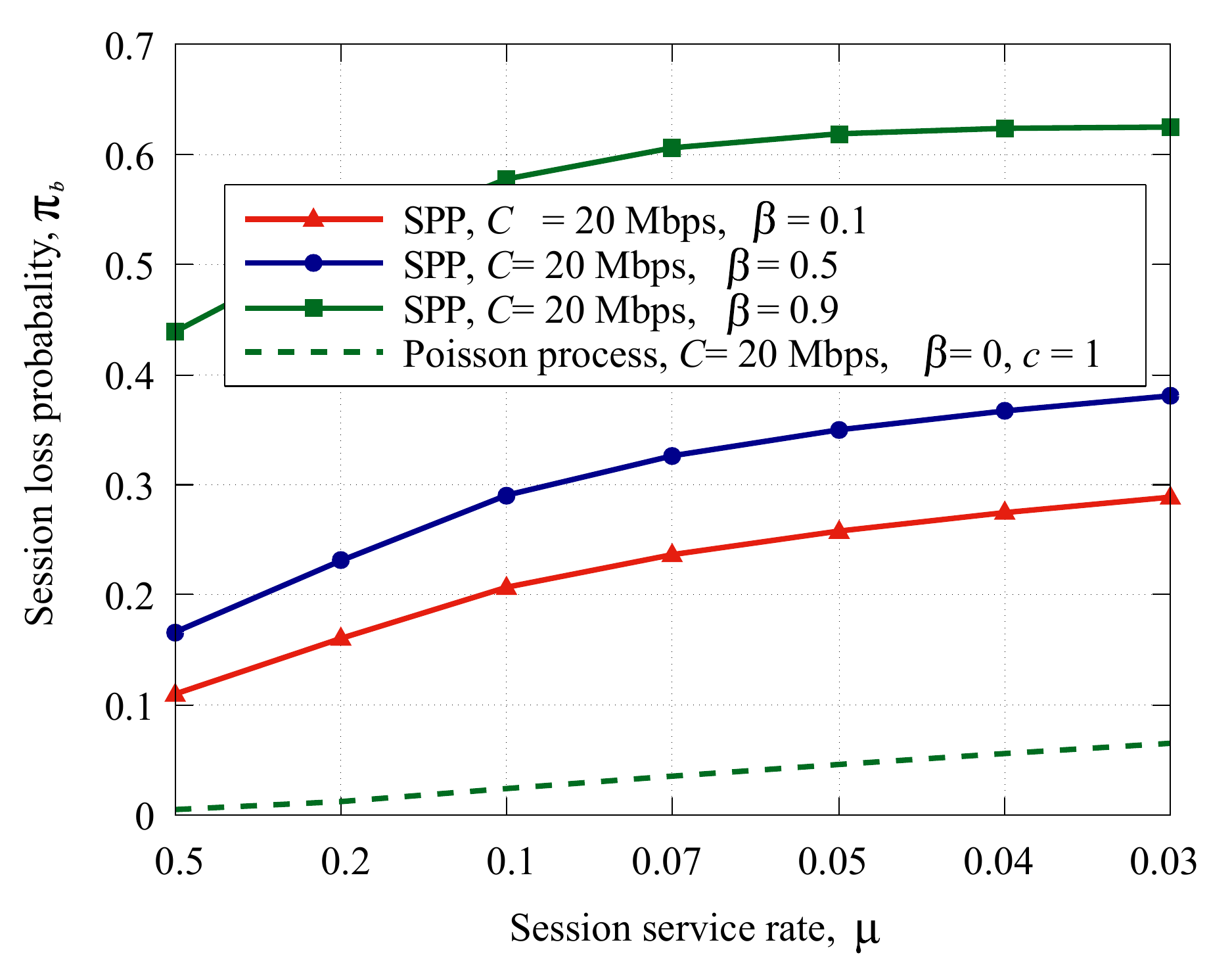}
    \label{fig:mu_loss}
}\vspace{-0mm}\\
\subfigure[{System resource utilization}]{
    \includegraphics[width=0.45\textwidth]{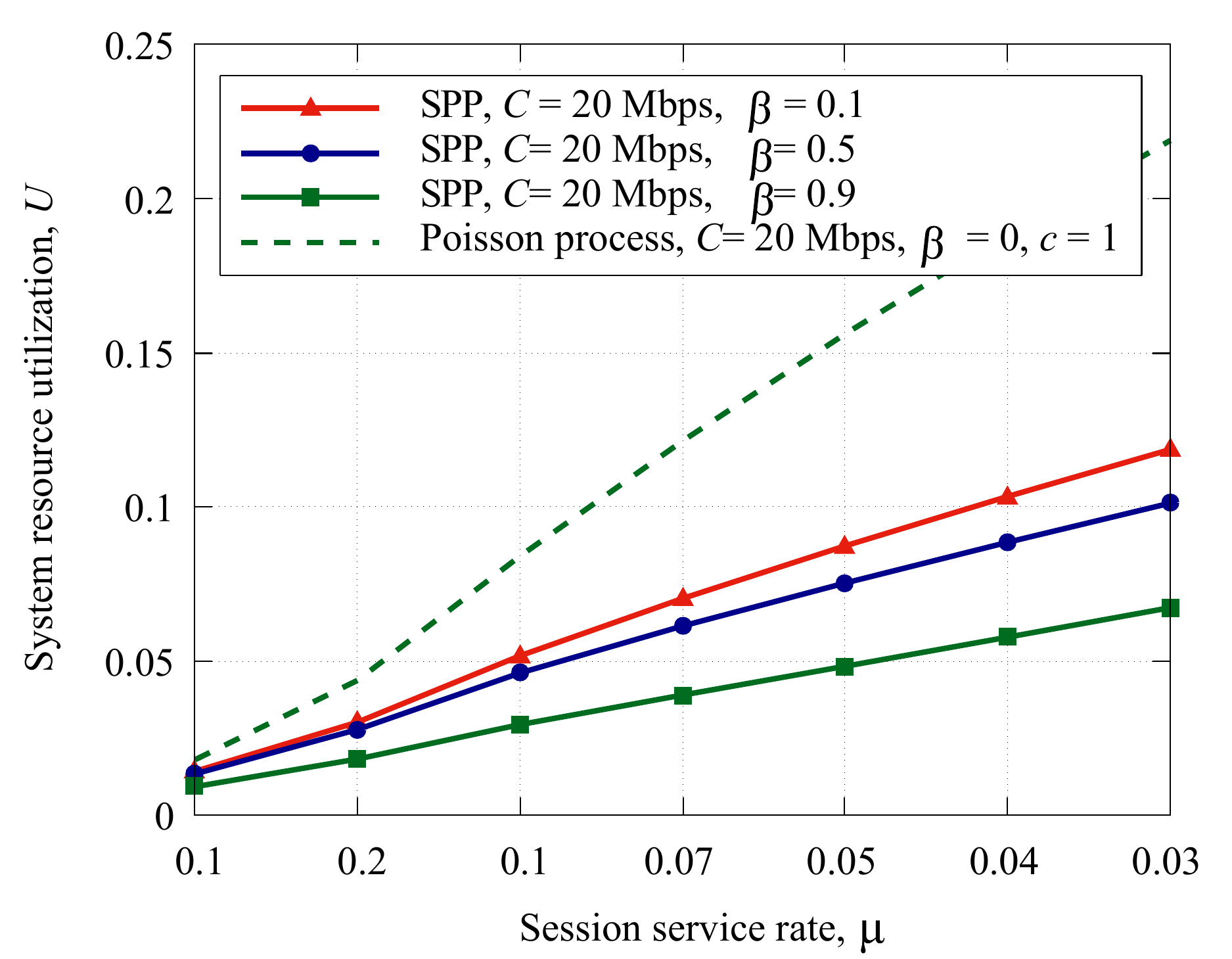}
    \label{fig:mu_util}
}
\vspace{-0mm}
\caption{Performance metrics as a function of the service rate.}
\label{fig:mu}
\vspace{-0mm}
\end{figure}


\textcolor{black}{We note that the increase of the CoV (Fig. \ref{fig:cov}) or lag-1 NACF (Fig. \ref{fig:beta}) results in the growth of burstiness of the arrival process, that is, the periods of low activity (session arrivals) and interchanged with periods of high activity for the same session arrival rate. Such behavior has been observed in classic Internet traffic \cite{jiang2005internet}, and in cellular systems as well \cite{zhang2012understanding,tsakmakis2020effect,jang2021resource}. It is known that burstiness of sessions arrivals cause significant drop in communication systems’ performance measures \cite{latouche1999introduction}. That is why, the behavior of the session loss probability and the resource utilization on Fig. \ref{fig:beta} and Fig. \ref{fig:cov} was expected.}


We note that the quantitative impact of lag-1 NACF, CoV, and PMF of the resource requirements depends heavily on other system parameters. As an example, Fig. \ref{fig:mu} shows the impact of the service rate $\mu$ on the session loss probability and system resource utilization for different values of lag-1 NACF value $\beta_A$, CoV $c_A=2$, session rate $C=20$ Mbps, BS antenna array of $16\times{}16$ elements, and session arrival rate $\lambda_A=0.1$ sess./s. Here, we see that different values of the lag-1 NACF lead to different user- and system-centric performance metrics for the same $\mu$ and this difference can be as high as an order of magnitude for session loss probability and up to two times for system resource utilization. 

\begin{figure}[!t]
\vspace{-0mm}
\centering\hspace{-0mm}
\subfigure[{Session loss probability}]{
    \includegraphics[width=0.45\textwidth]{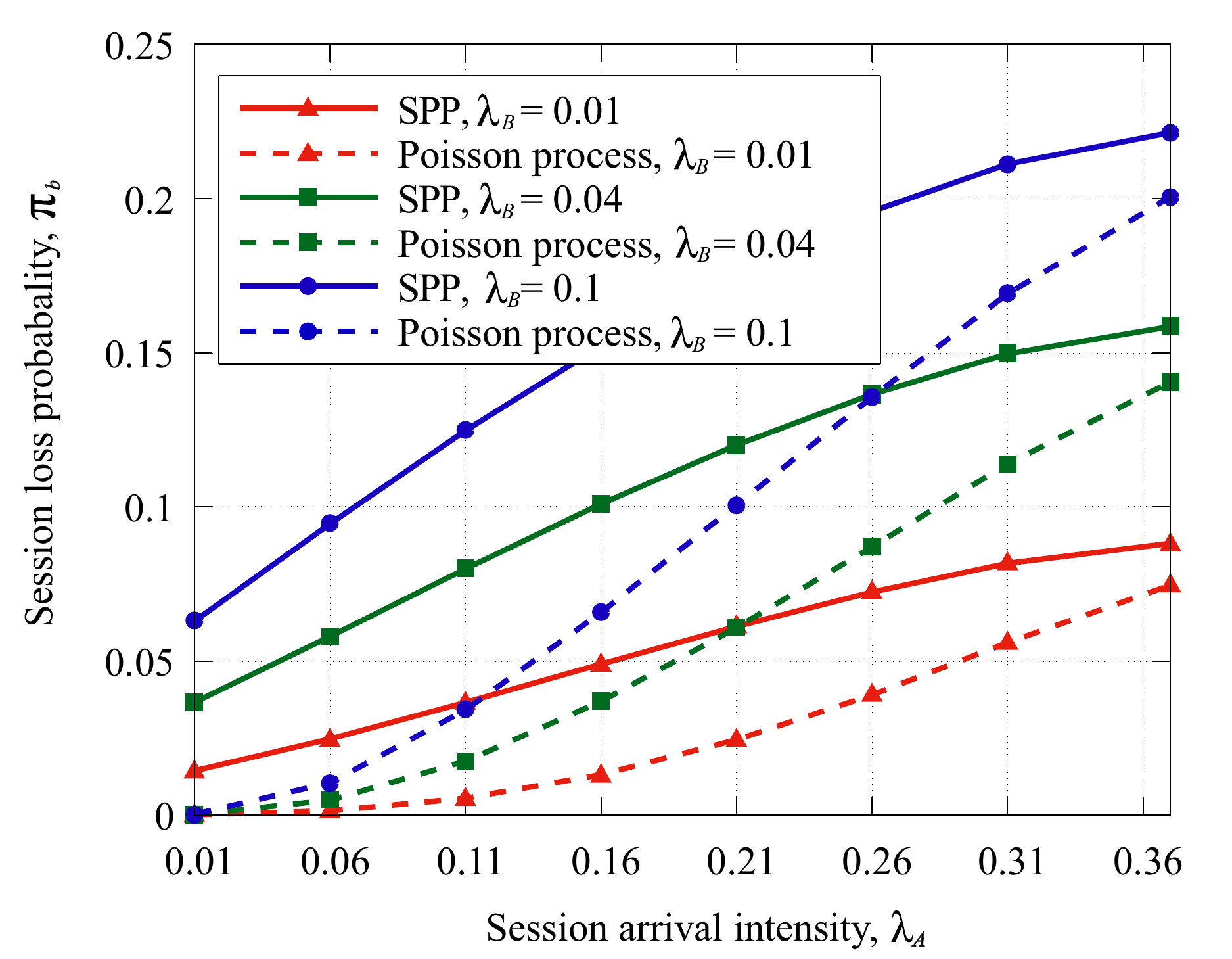}
    \label{fig:blockDens_loss}
}\vspace{-0mm}\\
\subfigure[{System resource utilization}]{
    \includegraphics[width=0.45\textwidth]{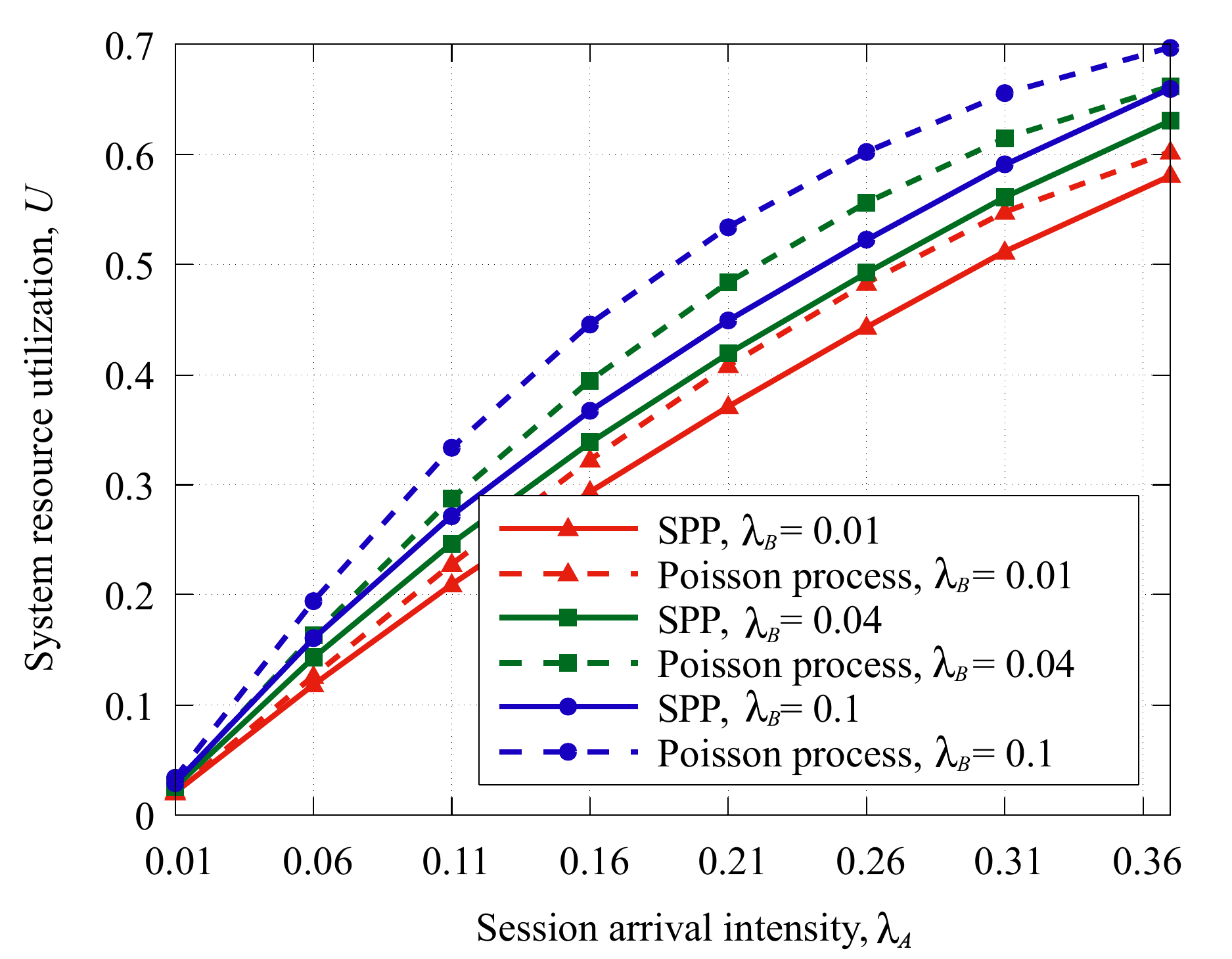}
    \label{fig:blockDens_util}
}
\vspace{-0mm}
\caption{The impact of the blockers' density.}
\label{fig:blockDens}
\vspace{-0mm}
\end{figure}

\subsection{Radio Part Parameters}


Finally, we would like to note that the radio part and environmental parameters, including antenna gains, emitted power, and blockers' density, also affect the magnitude of lag-1 NACF and CoV impact on session loss probability and system resource utilization. The impact of the first two parameters is mainly due to a change in the effective coverage radius of the BS captured in (\ref{eqn:coverage}). The larger the coverage radius the higher the variability of the PMF of the requested resource by a session and, thus, the more expressed the effects of the PMF of resource requirements, as demonstrated in Fig. \ref{fig:lambda_A}. 

Similarly, different values of blockers' density also lead to different variability in the PMF of resource requirements. Specifically, we demonstrate its impact on the session loss probability and system resource utilization in Fig. \ref{fig:blockDens}, where the CoV is $c_A=2$, session rate is $C=10$ Mbps, BS antenna array is $16\times{}16$ elements, session arrival rate is $\lambda_A=0.1$ sess./s, mean session service time is 30 s, and lag-1 NACF is $\beta_A=0.1$. By analyzing the presented data, we observe that the larger values of the blockers' density $\lambda_B$ lead to greater values of the session loss probabilities for both Poisson and correlated SPP arrivals. The rationale is that an increase in blockers' density increases the mean values of the amount of resources requested by a session. Specifically, the difference between the low blockage and high blockage scenarios can reach two times in terms of the considered parameter. However, the impact of blockers' density on system resource utilization is much milder, as shown in Fig. \ref{fig:blockDens_util}. Here, the maximum difference between the curves corresponding to the SPP and Poisson arrivals is approximately $0.05$. Thus, we may conclude that blockers' density mainly affects the user-centric performance metric.


\section{Conclusion}\label{sect:concl}


Motivated by the lack of studies on the impact of arrival process characteristics in 5G/6G mmWave/sub-THz systems on user- and system-centric performance measures, we developed a model of the service process of sessions arriving according to the MAP process with flexible statistical properties. The model accounts for the major specifics of mmWave/sub-THz systems including blockage, propagation, and antenna properties, and explicitly captures the evolution of the session service process at the BS. By utilizing a special case of the MAP process -- SPP process with an easily controllable CoV and lag-1 NACF, we investigated the impact of lag-1 NACF, CoV of the arrival process, and PMF of the session resource requirements on system- and user-centric performance measures, including session loss probability and system resource utilization.

\textcolor{black}{Our numerical results show that the autocorrelation function and CoV of the arrival process have a drastic impact on user- and system-centric performance metrics.} In deployments characterized by large values of these parameters, the performance is significantly worse than that of the completely uncorrelated Poisson session arrivals conventionally assumed in the performance analysis. The difference can reach an order of magnitude for the session loss probability and multiple times for the system resource utilization. Moreover, even a small deviation from the Poisson process assumption, that is, lag-1 NACF of $0.1$ and a CoV of $2$ already results in two times higher session loss probability and a decrease in system resource utilization by $10-30$\%. Finally, we demonstrated that the impact of the radio part parameters is mainly due to the increased variability of session resource requirements. Thus, by utilizing the Poisson arrival process as a session arrival process in 5G/6G mmWave/sub-THz systems, one can severely overestimate not only the provided quality of service parameters but also the efficiency of the system.  

\textcolor{black}{We note that the proposed model allows for extensions to more complex service processes. The extension to the case of multi-connectivity operation \cite{3gpp_MC} is feasible by following \cite{begishev2021joint,moltchanov2022tutorial} and the only limiting factor is the state-space that needs to be enlarged to account for dynamic handovers caused by the loss of connectivity between different serving cells during the session service. The results of the study in \cite{moltchanov2019analytical} can also be applied to account for dynamic user’s mobility within the cell. Essentially, this will also result in state space growth as the blockage process will follow Markov model. Inter-cell mobility, however, is more difficult to capture as this will induce the change of the sets of serving cells as user roams in the network. Such complex behavior should be captured by utilizing different methods, e.g., by system level simulations.}



\balance
\bibliographystyle{ieeetr}

\end{document}